\documentclass[a4paper]{JHEP3}
\pdfoutput=1 
\usepackage{amssymb}
\usepackage{amsmath}
\usepackage{cite}
\usepackage{graphicx}
\def\be{\begin{equation}}
\def\ee{\end{equation}}
\def\baray{\begin{eqnarray}}
\def\earay{\end{eqnarray}}
\def\ba{\begin{eqnarray}}
\def\ea{\end{eqnarray}}

\title{Signatures of Initial State Modifications on Bispectrum Statistics}

\author{P. Daniel Meerburg$^{1,2}$, Jan Pieter
van der Schaar$^{2,3}$ and Pier Stefano Corasaniti$^4$\\
$^1$ Astronomical Institute ``Anton Pannekoek", University of Amsterdam,\\
Kruislaan 403, 1098 SJ Amsterdam , The Netherlands\\
$^2$ Institute for Theoretical Physics, University of Amsterdam,\\
Valckeniersstraat 65, 1018 XE Amsterdam, The Netherlands \\
$^3$  Korteweg-de Vries Institute for Mathematics, University of Amsterdam,\\ 
Plantage Muidergracht 24 ,1018 TV Amsterdam, The Netherlands \\
$^4$ LUTH, Observatoire de Paris, CNRS UMR 8102, Universit\'e Paris Diderot, \\
5 Place Jules Janssen, 92195 Meudon Cedex, France\\}
\date{\today}

\abstract{Modifications of the initial-state of the inflaton field can induce
a departure from Gaussianity and leave a testable imprint on the higher order 
correlations of the CMB and large scale structures in the Universe. We focus on 
the bispectrum statistics of the primordial curvature perturbation
and its projection on the CMB. For a canonical single-field action the 
three-point correlator enhancement is localized, maximizing in the 
collinear limit, corresponding to enfolded or squashed triangles in 
comoving momentum space. We show that the available local and equilateral 
template are very insensitive to this localized enhancement and do not 
generate noteworthy constraints on initial-state modifications.
On the other hand, when considering the addition of a dimension $8$ 
higher order derivative term, we find a dominant rapidly oscillating 
contribution, which had previously been overlooked and whose significantly 
enhanced amplitude is independent of the triangle under consideration. 
Nevertheless, the oscillatory nature of (the sign of) the correlation function
implies the signal is nearly orthogonal to currently available
observational templates, strongly reducing the sensitivity to the enhancement. 
Constraints on departures from the standard Bunch-Davies vacuum state 
can be derived, but also depend on the next-to-leading terms.
We emphasize that the construction and application of especially 
adapted templates could lead to CMB bispectrum constraints on modified initial 
states already competing with those derived from the power spectrum.}

\keywords{Inflation, CMB, non-Gaussianities}

\preprint{ITFA-2008-53}

\begin{document}

\section{Introduction}

It has been known for a while now \cite{Kofman1991, Komatsu2002}
that the potential presence of non-Gaussian signatures in the CMB is a powerful probe of 
the physics of inflation and beyond. Computations
of the primordial bispectra \cite{ghostinfl_ng, exotic_ng, curvaton_ng, 
single_field_ng, inflation_ng, multiefield_ng, ekpyrosis_ng, Bartolo2004, 
single_field_ng_2} (and later trispectra \cite{Komatsu2002, trispectrum_ng}) have shown that
different models of inflation can produce rather unique features,
which would allow, when detected, to discriminate between them.
For the bispectrum, the distinction between models relies
on two features 1) the overall amplitude of the non-Gaussian signal
and 2) the detailed dependence on the comoving momenta. Obviously,
when the overall amplitude of the signal is low, the second feature will
be much harder to observe. Observational limitations due to foreground
contamination \cite{Creminelli2004, Serra2008} and cosmic variance 
limit the detection of non-Gaussianity in the CMB temperature and polarization 
spectrum \cite{Komatsu2001a, Babich2004b}. For that reason one can already 
conclude that non-Gaussianity should be observably absent if a single, slowly rolling,
scalar field is responsible for inflation \cite{Riotto2002, Maldacena2002}. 

Even if a model predicts a detectable non-Gaussian amplitude,
it will remain a challenge to measure the actual momentum
dependence, since the inferred constraints on the level of non-Gaussianity
\cite{Komatsu2003a, Creminelli2005, CSZT2007, Yadav2007c, Komatsu2008} 
are based on a sum over all modes of a pre-assumed momentum dependence\footnote{Other methods such as Minkowski Functionals (see \cite{Schmalzing1995} for theory and \cite{Komatsu2008} and references therein for observational results) and a Wavelet approach (\cite{Hobson1999} and subsequent papers) exist which typically do not rely on a pre-assumed momentum dependence. Here however we refer to the approach initiated in \cite{Komatsu2001a} and further developed in \cite{Komatsu2003,Babich2004a,Liguori2005,Kogo2006,Creminelli2006,Yadav2007a,Liguori2007,Yadav2007b}, which seems to give the most consistent and stringent constraints \cite{Komatsu2008} so far.}. 
Such dependencies are known as `local' (or `squeezed') 
and `equilateral' template, which correspond to particular shapes that maximize in some `extreme' 
triangle configuration in momentum space. The possibility to distinguish between different 
theoretical models producing a sizable non-Gaussian amplitude relies on the fact that in the 
models considered so far the produced non-Gaussianities are well approximated by {\it one} of these 
templates. For example, it has been shown \cite{single_field_ng_2, Khoury:2008wj} that non-canonical
kinetic terms and higher derivative contributions to the inflaton potential can produce
significant levels of non-Gaussianity of the equilateral type if the speed of sound in these 
models is much smaller than the speed of light, which can be realized in 
certain brane inflation scenarios \cite{Silverstein:2003, Chen:2004}.
Local shape non-Gaussianities were the first type to be considered \cite{fnl_local, Pyne1995, Komatsu2001a} 
and are a direct consequence of the nonlinear relation between the inflaton fluctuations and the curvature 
perturbations that couple to matter and radiation. In \cite{Riotto2002, Maldacena2002} it was shown that the 
amplitude of local type non-Gaussianities in single-field slow-roll inflation is proportional 
to the slow-roll parameter $\epsilon$ \footnote{This is the semi-classical, tree level, result. 
The effects of quantum loop corrections have been studied in \cite{loop_corrections}.}, which is very small by construction. The amplitude of local type 
non-Gaussianities should therefore be undetectably small if single-field slow-roll is responsible 
for inflation. In contrast, large local non-Gaussianities can be generated in curvaton 
models \cite{curvaton_ng}, where the curvature perturbation $\zeta$ can evolve outside the horizon, 
or inflationary models with multiple scalar fields. Models of new ekpyrosis \cite{ekpyrosis_ng} and the recently proposed 
contracting models with an increasing speed of sound \cite{Khoury:2008wj} , in which a bouncing 
universe is replacing inflation, also yield large non-Gaussianities of the local type.  

In this paper we will focus on non-Gaussian features arising from an arbitrary initial-state modification. 
This type of non-Gaussianity has been discussed in \cite{Martin1999,Gangui2002,Porrati2004a, Porrati2004b, Shiu2006, Holman2007}, 
here we provide a more detailed analysis on their detectability.
In the language of boundary effective field theory \cite{SSS2004} one can generally divide the contributions into two parts; 
non-Gaussianities coming directly from the initial-state boundary (which are absent when considering Gaussian initial-state modifications), 
and `bulk' non-Gaussianities generated by the presence of (interacting) particles in the modified initial-state
\cite{Meerburg2008}. In the boundary effective field theory formalism it has been shown that the leading non-Gaussian 
initial-state modification is of the local type \cite{Porrati2004a, Porrati2004b}. However, the `bulk' 
non-Gaussianities generated by the non-zero Bogoliubov coefficient seem to have a unique momentum dependence,
which is very different from that of the local and equilateral types \cite{single_field_ng_2, Holman2007,
Meerburg2008}. For example, for a canonical single-field inflaton action, in momentum space the non-Gaussian signal 
produced by a, possibly Gaussian, modification of the initial-state maximizes for triangles where two momentum vectors are collinear, 
i.e. when the magnitude of one of the comoving momenta equals the sum of the other two: $k_p=k_q+k_r$ with $p\neq q\neq r$ 
(a squashed, flattened or enfolded triangle), and are known as collinear or enfolded type non-Gaussianities. 

Thus far only the local and equilateral type of non-Gaussianities have been constrained by 
the data \cite{CSZT2007, Komatsu2008}, although recently a strong case has been made for a more 
general set-up \cite{Fergusson2008}. There are essentially two reasons for this. 
First of all, the realization that initial-state modifications give rise to a unique non-Gaussian shape, 
that might even be detectable due to subtle enhancements, is rather recent and its theoretical motivation 
might be considered less compelling. Putting aside plausible theoretical concerns associated with
modifications of the vacuum state and instead taking a phenomenological point of view, deviations from the 
standard Bunch-Davies state are tightly constrained \cite{Spergel2006} because of their unique oscillatory
signatures in the $2$-point power spectrum (see \cite{GSSS2005} and references therein). 
As emphasized in \cite{Holman2007} the bispectrum (and possibly even higher $n$-point functions) 
might be as good, or even better, in constraining initial-state modifications. Clearly, with the
expected future improvements in detecting primordial non-Gaussian signals, it is 
worthwhile to look for the presence of enfolded type or, as we will argue in this paper, oscillatory 
non-Gaussianities in the CMB data to constrain initial-state modifications.
The second more pragmatic reason why enfolded type non-Gaussianities have not been compared to the data yet 
is that in analyzing the data computational limitations demand that the momentum dependence is 
{\it factorizable}. Generic $3$-point correlators are not factorizable, so one resorts to constructing a 
factorizable template that approximates the actual theoretical bispectrum, maximizing in the 
appropriate `extreme' triangle. It is this factorizable template that is then compared to the data. 
Such templates have been constructed for the local \cite{Komatsu2002} and equilateral shapes 
\cite{Creminelli2005}, but has not yet been constructed for the type of non-Gaussianities predicted by 
initial-state modifications, which are typically expected to extremize in an enfolded (collinear, squashed 
or flattened) triangle. The goal of this paper is two-fold: to present a detailed analysis on the 
detectability of non-Gaussianities produced by initial-state modifications using currently available templates, 
both with and without higher derivative corrections, and secondly to determine how much improvement 
can theoretically be gained by using more optimal templates.

The paper is organized as as follows. In Section~\ref{Preliminaries} we review the standard analytical tools to 
study non-Gaussianity, in particular the computation of the 3-point correlation function in momentum space 
and its relation to different triangular shapes. In Section~\ref{SingleField} we will present a 
detailed analysis of the 3-dimensional bispectrum from initial-state modifications in the 
single-field slow-roll inflationary scenario. In Section~\ref{HD} we analyze the case of 
modified initial-state non-Gaussianities in the presence of a dimension $8$ higher order derivative term 
in the Lagrangian. In Section~\ref{CMB} we discuss the results of the CMB bispectrum
computation and finally we present our conclusions in Section~\ref{Conclusion}.

\section{Three-dimensional bispectrum preliminaries}\label{Preliminaries}

In this section we will briefly review the standard tools for analyzing non-Gaussianities 
as first described in \cite{Babich2004a}. In the next sections we will apply these tools to
the case of initial-state modifications. Let us start considering the primordial spectrum of 
curvature perturbations generated by the inflaton. In three-dimensional comoving momentum space 
a generic three-point correlator of the curvature perturbation $\zeta_{\vec{k}}$ 
is a function of the three comoving momenta $\vec{k}_1$, $\vec{k}_2$ and $\vec{k}_3$, 
which in $3$ dimensions corresponds to a total of 9 parameters. 
Translational invariance forces the three-point function 
to conserve momentum
\begin{eqnarray}
\langle\zeta_{\vec{k}_{1}}\zeta_{\vec{k}_{2}}\zeta_{\vec{k}_{2}}\rangle & = & A \cdot (2\pi)^{3}\delta\left(\sum_i \vec{k}_{i} \right) F(\vec{k}_{1},\vec{k}_{2},\vec{k}_{3}), \label{eq:translationalinvariance}
\end{eqnarray} 
which fixes one of the momenta, reducing the number of free parameters from $9$ to $6$. 
Rotational invariance allows one to pick a $2$-dimensional plane defined by the remaining two momenta and 
adjust the axes such that one of the momenta is along one of the axes of the plane. This fixes another $2+1=3$ 
parameters, leaving only 3 variables to parametrize the three-point correlator. These can be identified with
two angles and the overall scale of the triangle formed by $\vec{k}_1$, $\vec{k}_2$, $\vec{k}_3$. 
Since the primordial power spectrum is approximately scale invariant, 
we expect the correlator to be a homogeneous function $F$ of degree $-6$ in comoving momentum space, i.e.
$F(\lambda \vec{k}_1, \lambda \vec{k}_2, \lambda \vec{k}_3)= \lambda^{-6} F(\vec{k}_1, \vec{k}_2, \vec{k}_3)$. 
So (approximate) scale invariance fixes the dependency of the three-point correlator on the scale of the 
triangle, further reducing the number of free parameters to the $2$ angles. Instead of writing the function 
$F$ in terms of these angles, it is most convenient to consider the two independent ratios given by 
the magnitudes of the comoving momenta $x_2 \equiv k_2/k_1$ and $x_3 \equiv k_3/k_1$. 
In order to determine the relevant $x_2$, $x_3$ domain, 
one assumes $k_1 \geq k_2 \geq k_1$, giving $x_2 \leq 1$ and $x_3 \leq 1$,  and then uses the triangle constraint 
to find that $1-x_2 \leq x_3 \leq 1$, identifying the top-right triangle in $x_2$, $x_3$ space (see e.g. figure \ref{fig:enfoldedtemplate}). 
Since the distributions are symmetric in $x_2$ an $x_3$, one could further reduce the domain by half 
only considering $x_3 \geq x_2$. Hitherto, unlike the power spectrum, which only 
depends on the `reciprocal distance' between two-points, the
bispectrum $F$ depends on two variables, typically represented by the ratios of the magnitudes of the 
comoving momenta $F=F(x_2,x_3)$. 

To measure the overall amplitude $A$ in Eq.~(\ref{eq:translationalinvariance}), 
one assumes a particular theoretical template shape function $F(x_2,x_3)$, 
sums over all triangles and then normalizes appropriately, taking into account 
the variance of a given mode in Fourier space. This procedure leads to the following
estimator of the non-Gaussian amplitude $A$
\begin{eqnarray}
\hat{A} & = & \frac{ \sum_{\vec{k}_i} \zeta_{\vec{k}_{1}}\zeta_{\vec{k}_{2}}\zeta_{\vec{k}_{2}} F(\vec{k}_1, \vec{k}_2, \vec{k}_3) / \left( \sigma^{2}_{k_1}\sigma^{2}_{k_2}\sigma^{2}_{k_3}\right)}{\sum_{\vec{k}_i} F^2(\vec{k}_1, \vec{k}_2, \vec{k}_3) / \left( \sigma^{2}_{k_1}\sigma^{2}_{k_2}\sigma^{2}_{k_3}\right)}
.\label{bestestimator}
\end{eqnarray}
Here the $\sigma_k$ represent the variances of the different modes and the sum runs over all 
triangles in momentum space. The above estimator naturally defines a 
scalar product between two distributions $F_X$ and $F_Y$ as \cite{Babich2004a}
\begin{eqnarray}
F_X\cdot F_Y & = & \sum_{\vec{k}_i} \frac{F_X(\vec{k}_1, \vec{k}_2, \vec{k}_3) F_Y(\vec{k}_1, \vec{k}_2, \vec{k}_3)}{\sigma^{2}_{k_1}\sigma^{2}_{k_2}\sigma^{2}_{k_3}}.\label{eq:shapeproduct}
\end{eqnarray}
This scalar product allows us to quantitatively verify how 
well a particular template distribution, say $F_X$, can be used to constrain a theoretical signal described 
by the distribution $F_Y$. In terms of the (reduced set of) parameters $x_2, x_3$ the sum over 
triangles can be written as an integral with an appropriate measure equal to $x_2^4 x_3^4$ 
\begin{eqnarray}
F_X\cdot F_Y &\propto&\int d x_2 dx_3 F_X(x_2,x_3)F_Y(x_2,x_3)x_2^4x_3^4.\label{eq:scalarproduct}
\end{eqnarray}
To derive optimal constraints using a template $F_X$ one would like the scalar product, or the overlap, 
to be as large as possible. Using  the scalar product one can construct a normalization independent `cosine' 
between two distributions
\begin{eqnarray}
\mathrm{Cos}(F_X,F_Y)&\equiv& \frac{F_X\cdot F_Y}{(F_X\cdot F_X)^{1/2} (F_Y\cdot F_Y)^{1/2}} \, ,
\label{eq:cosine}
\end{eqnarray}
which is close to $1$ for shapes that are very similar and considerably smaller than $1$
for shapes that are very distinct. It follows that optimal constraints can be obtained 
only if the (factorizable) templates, which are used to analyze the data, have a cosine 
close to $1$ with the theoretically predicted non-Gaussian signal. Nevertheless, for non-optimal 
templates $F_X$ one can still derive constraints on a theoretically predicted non-Gaussian signal $F_Y$
provided one introduces the so-called fudge factor $\Delta_F$, defined as 
\cite{Babich2004a}
\begin{eqnarray}
\Delta_F &=& \frac{F_Y \cdot F_X}{(F_X \cdot F_X)} = \mathrm{Cos}(F_X,F_Y) 
\left( \frac{F_Y \cdot F_Y}{F_X \cdot F_X}\right)^{1/2}  \, .
\label{eq:fudgefactor}
\end{eqnarray}
The fudge factor allows to deduce the relevant constraints for different theoretical predictions $F_Y$ 
using the results inferred from the data analysis of a particular template distribution $F_X$. 
In such a case the constraint on the amplitude of the type $F_X$ will 
be degraded by a factor $1/\Delta_F$, thus the smaller the scalar product between $F_Y$ and $F_X$ the
weaker the constraints on the $F_Y$ type non-Gaussianities using the $F_X$ template. 
Looking at Eq.~(\ref{eq:fudgefactor}) it should be clear that optimal constraints can be achieved by maximizing the cosine between
the template and the theoretical prediction. The other contribution to the fudge factor has to do with some conventional choice 
of normalization for the template and the theoretical distribution involved and can be adapted accordingly. 
We will apply these techniques to obtain constraints on non-Gaussianities 
predicted by modified initial-states using the latest results on local and equilateral type non-Gaussianities,
and to derive what can (theoretically) be gained by analyzing the data with an more optimal (enfolded) 
template.

Let us briefly discuss the normalization conventions for the non-Gaussian amplitudes, 
which are important for a correct interpretation and comparison of the results obtained
for different distributions. 
To compare the local and equilateral template one typically equates the 
distributions in the equilateral triangle 
$k_1=k_2=k_3$ \cite{Babich2004a}. We will follow this convention, which allows 
us to directly use the constraints from the CMB for the local and equilateral non-Gaussian amplitudes. 
To be explicit, for the local template distribution the 
standard definition of the $f_{\mathrm{NL}}^{\mathrm{local}}$ parameter, starting from the general three-point function 
in Eq.~(\ref{eq:translationalinvariance}), is related to the amplitude $A$ of the three-point 
function of curvature perturbations in the following way\footnote{Our sign convention for the non-Gaussian amplitude follows 
\cite{Babich2004a,Maldacena2002}, which is different from that used in \cite{Komatsu2003}.} 
\begin{eqnarray}
A  &=& (2\pi)^4 \left(-\frac{3}{5}f_{\mathrm{NL}}^{\mathrm{local}}\right) \, 
\frac{\Delta_{\Phi}^{2}}{k_1^6} \, , \label{eq:fnlocal}
\end{eqnarray}
where $\Delta_{\Phi} = \frac{1}{8\pi^2} \frac{H^2}{\epsilon M_{p}^2}$  is the amplitude of the 
two-point power spectrum, which has been observed to be approximately equal to $10^{-10}$, 
$M_p$ is the reduced Planck mass and $\epsilon=\frac{1}{2} M_p^2 \left( \frac{V'}{V} \right)^2$ 
is the first slow-roll parameter. In the above expression for the amplitude $A$
we included the overall $k_1$ scaling dependence, implying that the local shape $F^{\mathrm{local}}$ 
can be identified as the following function of the reduced number of variables $x_2$, $x_3$
\begin{eqnarray}
F^{\mathrm{local}}(x_2,x_3)  &=& 2 \left(\frac{1}{x_{2}^{3}}+\frac{1}{x_{3}^{3}}+\frac{1}
{x_{2}^{3} x_{3}^{3}}\right) \, , \label{eq:localshape}
\end{eqnarray}
For non-Gaussianities of the equilateral type it was shown in \cite{Creminelli2005, Babich2004a} 
that these are well approximated by the following shape function
\begin{eqnarray}
F^{\mathrm{equil}}(x_{2},x_{3}) & = &  6 \left[-\frac{1}{x_{2}^{3}}-\frac{1}{x_3^3}-\frac{1}{x_2^3 x_3^3}
-\frac{2}{x_{2}^{2}x_{3}^{2}} +\left(\frac{1}{x_{2}^{2}x_{3}^{3}}+5\;
\mathrm{perm}\right)\right], \label{eq:equilshape}
\end{eqnarray}
where the normalization has been fixed such that the local and equilateral template shape functions 
both equal $6$ in the equilateral limit $x_2 =x_3=1$. Comparing to the local template definition of
the non-Gaussian amplitude $f_{\mathrm{NL}}$, this then suggests a similar definition of 
$f_{\mathrm{NL}}^{\mathrm{equil}}$ 
\begin{eqnarray}
A  & = & (2\pi)^4 \left( -\frac{3}{5}f_{\mathrm{NL}}^{\mathrm{equil}}\right) \, \frac{\Delta_{\Phi}^{2}}{k_1^6}.\label{eq:fnlequil}
\end{eqnarray}
A crucial property of the local and equilateral template is that they are factorized in 
their comoving momentum dependence. This allows for a drastic (and necessary) reduction in the 
computational time needed to compare the template distributions to the CMB data, 
yielding constraints on the parameters $f_{\mathrm{NL}}^{\mathrm{local}}$ and 
$f_{\mathrm{NL}}^{\mathrm{equil}}$. The analysis of the WMAP-5 year data
for local and equilateral non-Gaussianities gives the following limits, 
\cite{Komatsu2008}
\begin{eqnarray}
-9 &<& f_{\mathrm{NL}}^{\mathrm{local}} < 111 \nonumber \\
-151 &<& f_{\mathrm{NL}}^{\mathrm{equil}} < 253 \, ,
\end{eqnarray}
which we will use in Section~\ref{CMB}.
It is worth stressing that the non-Gaussian amplitude $f_{\mathrm{NL}}$
is not uniquely defined, it depends on a specific choice for the shape function $F_X$, which is equivalent to fixing the integrated 
norm $|F_X| \equiv \sqrt{F_X \cdot F_X}$. It is the combination $f^X_{\mathrm{NL}} \, |F_X| $ that is independent of a particular 
normalization scheme and which measures the (integrated) non-Gaussian amplitude. Obviously any choice will do, as long
as one properly takes into account the corresponding norm $|F_X|$ when for example deducing constraints on the non-Gaussian 
amplitude $f_{\mathrm{NL}}^X$  from the equilateral and local template results. 

\section{Modified initial-state non-Gaussianities}\label{SingleField}

Theoretically predicted three-point functions, evaluated in the regular Bunch-Davies
vacuum state, describe a non-Gaussian signal of either local or equilateral type, 
depending on whether higher derivative corrections play a significant role in the 
inflationary evolution. If this is the case, as in DBI models of inflation \cite{Silverstein:2003, Silverstein:2004}, 
then the dominant contribution is of the equilateral type and can be large enough to be 
detectable in the near future. The existence of different shapes 
can be nicely understood in terms of the nonlinear origin of the 
non-Gaussian signal. For the local shape it is the nonlinear relation between the inflaton and the 
curvature perturbation on super-horizon scales that produces the maximal effect, whereas 
in the DBI case nonlinear effects in the inflaton sector are most relevant and maximize 
when all momenta cross the horizon. 

As was shown in \cite{Shiu2006, Holman2007} non-Gaussian effects can also be generated by 
dropping the assumption that the vacuum state is Bunch-Davies. To fundamentally address the vacuum state
ambiguity one would first need a full understanding of physics at the highest energy scales,  where the description 
in terms of a free inflaton field is expected to break down, as well as the physics preceding 
inflation. Nevertheless departures from the free Bunch-Davies state can be studied on a phenomenological
basis and it seems worthwhile to use observations to constrain the possibilities. The two-point power 
spectrum already provides strong constraints on the initial-state, which has to be close to Bunch-Davies 
\cite{Spergel2006}. Interestingly though, according to \cite{Holman2007}, three-and higher $n$-point functions 
might be very constraining as well, mainly due to subtle enhancement effects, which increase the non-Gaussian 
amplitude in collinear or enfolded triangles. In this section we will focus on the simplest case, with a 
three-point correlation function derived in the general context of  slow-roll inflation, but  
evaluated in a vacuum state different from standard Bunch-Davies, as parametrized by an 
undetermined Bogoliubov parameter $\beta_k$. The leading non-Gaussian contribution due to the 
appearance of a negative frequency mode is essentially obtained by swapping the sign of one
of the comoving momenta in the slow-roll inflation result \cite{Maldacena2002}. In appendix B we confirm 
the result first derived in \cite{Holman2007} that the correction to the three-point correlation function is given by 
\begin{equation}
\langle\zeta_{k_{1}}\zeta_{k_{2}}\zeta_{k_{3}}\rangle^{\mathrm{nBD}}=(2\pi)^{3}\delta^{(3)}\left(\sum\vec{k}_{i}\right)\frac{1}{M_{p}^{2}}\frac{4}{\prod(2k_{i}^{3})}\frac{H^{6}}{\dot{\phi}^{2}}\sum_{j}\frac{3k_{1}^{2}k_{2}^{2}k_{3}^{2}}{k_{j}^{2}\tilde{k}_{j}}\mathcal{R}e(\beta_{k_{j}})\left(\mathrm{cos}(\tilde{k}_{j}\eta_{0})-1\right)\label{modin2}
\end{equation}
In the above expression $\tilde{k}_j = \sum_i k_i - 2 k_j$ and $\eta_0$ represents the initial 
conformal time which has to be introduced to ensure that the non-Gaussian effects can consistently  
be calculated using an effective field theory description valid below some physical
cut-off momentum scale $M$ \cite{Holman2007}, i.e. non-Gaussianities 
generated by initial-state modifications are sensitive to the details of the ultraviolet complete 
theory and the physical cut-off scale $M$ is introduced to parameterize our ignorance. 
This suggests that the initial time $\eta_0$ should be a function of the comoving momentum $k$, 
allowing the combination $|k \, \eta_0(k)| = M/H \gg 1$ to be a large fixed number independent of $k$.
This prescription treats all comoving momenta equivalently, tracing back different comoving momenta from the time their 
physical momentum equals the cut-off scale $M$, preserving scale invariance. Instead considering $\eta_0$ 
to be some fixed initial time would immediately result in a breaking of scale invariance because different comoving momenta 
would receive contributions from a different range of physical momentum scales in the time integral involving the interaction 
Hamiltonian (see appendix B).

This is very reminiscent of the distinction between the New Physics Hypersurface (NPH) \cite{Danielsson2002, Easther2002} and
Boundary Effective Field Theory (BEFT) \cite{SSS2004, GSSS2005} proposals to model initial-state modifications. 
In the latter case one fixes an initial time where one calculates corrections to the usual Bunch-Davies initial condition 
using boundary effective field theory. The result is a Bogoliubov parameter $\beta_k$ depending on $k$, resulting in an explicit
breaking of scale-invariance in the two-point power spectrum. In the NPH scenario one traces every momentum mode back to some 
large physical cut-off scale $M$ and imposes the standard flat space vacuum state (corresponding to positive frequency modes only), 
mode by mode, resulting in a prediction for $\beta_k$ that is independent of $k$, which only gives rise to a small departure from 
scale-invariance after taking into account the slow-roll evolution of the Hubble parameter. Note that the NPH vacuum state proposal is not  
grounded (yet) in some effective field theory scheme that can systematically be applied to calculate quantum state corrections, as opposed 
to the BEFT approach. Both proposals have their problems and, as emphasized before, should at this stage and for our purposes here 
be considered as purely phenomenological models distinguished primarily by their consequences for scale-invariance. 

For the bispectrum, considering an initial time $\eta_0$ independent of $k$ (BEFT) or a $k$-independent combination $|k \eta_0(k)| = M/H$ (NPH) immediately results in a breaking of scale-invariance, even independent of any specific $k$-dependent prediction for $\beta_k$. Because the tools to analyze non-Gaussian shapes introduced in the previous section crucially rely on scale-invariance of the bispectrum, we will assume that the initial time $\eta_0(k)$ depends on the comoving momentum such that $|k \eta_0|=M/H$ is $k$-independent, in the spirit of the scale-invariant 
NPH scenario. It would be interesting to relax this assumption and apply more general techniques, for instance those recently developed in 
\cite{Fergusson2008}, to analyze the scale-dependent non-Gaussianities that arise in a BEFT approach of initial-state modifications.

Looking at the dependence on comoving momenta, we can see that the 
three-point correlator Eq.~(\ref{modin2}) maximizes in the collinear or enfolded triangle defined by $\tilde{k}_1/k_1 = \pi/ |k_1 \eta_0| \sim 0$, 
and with an amplitude proportional to $k_1 \eta_0$. Here $k_1$ is assumed to be the largest comoving momentum 
in the triangle, whose overall power-law dependence manifests the expected scale invariance of the 
three-point function. Similarly to the local and equilateral shapes, this enfolded type of non-Gaussianity 
can be associated to a dominant source of nonlinearities, in this specific case this is 
related to the unavoidable presence of interacting particles in the modified initial-state at sub-horizon scales. 
Based on the $|k_1 \eta_0|=M/H \gg 1$ enhancement 
of the non-Gaussian signal a rough order of magnitude estimate of the observational constraints on modified initial-state 
non-Gaussianities was given in \cite{Holman2007}. However, their estimate was inferred by considering the maximum signal 
in the enfolded limit, and directly compared to existing bounds on the local non-Gaussian amplitude. In contrast,
a full analysis of the sensitivity of current non-Gaussian constraints on departures from the Bunch-Davies 
vacuum must involve integrating over all triangles and crucially relies on the scalar product between 
the theoretical template prediction and the different observational template distributions. We will address this issue
throughout the rest of this paper.

In order to proceed and calculate the scalar product, cosine and fudge factor, we need to determine the dominant 
contribution to the shape function and identify the corresponding non-Gaussian amplitude. 
Starting from Eq.~(\ref{modin2}) we identify the relevant comoving momentum dependent part as
\begin{eqnarray}
F^{\mathrm{modin}}(k_{1},k_{2},k_{3}) & = & \frac{1}{k_1 k_2 k_3}\left\{ \frac{1-\cos [\eta_0(k_1+k_2-k_3)]}{k_3^2(k_1+k_2-k_3) }+ \frac{1-\cos [\eta_0(k_1-k_2+k_3)]}{k_2^2 (k_1-k_2+k_3)} + \right. \nonumber \\
& & \left. \frac{1-\cos [\eta_0(-k_1+k_2+k_3)]}{k_1^2(-k_1+k_2+k_3)}\right\} \label{fmodin},
\end{eqnarray}
which by having scaled out the standard $k_1^{-6}$ dependence leads to the corresponding definition for the amplitude $A$,
\begin{eqnarray}
A &=& (2\pi)^4 \, (3 \epsilon |\beta| ) \, 
\frac{\Delta_{\Phi}^{2}}{k_1^6}, \nonumber \label{eq:modinamplitude}
\end{eqnarray}
where we have replaced $\mathcal{R}e(\beta_k)$ with the absolute value $|\beta|$.
By comparing Eq.~(\ref{eq:modinamplitude}) to the amplitudes of the local and equilateral templates, a
standard definition of $f_{\mathrm{NL}}^{\mathrm{enf}}$ suggests that $f_{\mathrm{NL}}^{\mathrm{enf}}=5 \epsilon
|\beta|$. Consequently, without any enhancement from a large fudge factor this non-Gaussian amplitude is obviously 
undetectable, since it is suppressed by both the slow-roll parameter $\epsilon$ and the Bogoliubov parameter $|\beta|$.
In terms of the reduced variables $x_2$, $x_3$ we have
\begin{eqnarray}
F^\mathrm{modin}(k_1\eta_0,x_2,x_3)&=&\frac{1}{ x_2 x_3}\left\{\frac{1-\cos [k_1\eta_0(1+x_2-x_3)]}{x_3^2(1+x_2-x_3) }+\frac{1-\cos [k_1\eta_0(1-x_2+x_3)]}{x_2^2 (1-x_2+x_3)}+\right.\nonumber \\
& &\left.\frac{1-\cos [k_1\eta_0(-1+x_2+x_3)]}{(-1+x_2+x_3)}\right\}.\label{eq:modinshape2}
\end{eqnarray}
We now explicitly see the dependence of the shape function on $|k_1 \eta_0|=\frac{(|k_1/a_0|)}{H}=M/H$, 
namely the ratio of the physical cut-off scale to the Hubble parameter, as was explained earlier. Again, not 
fixing the combination $|k_1 \eta_0|$ to be $k_1$-independent results in an obvious breaking of 
scale-invariance and would not allow us to use the introduced tools for comparison with the available templates. 
The cut-off scale should be significantly larger than the Hubble scale and we 
will typically consider it to be somewhere in between $10^2 - 10^3$. This implies the shape function 
is rapidly oscillating, which complicates the evaluation of the integrals to determine 
the cosines and fudge factors with the available templates. When possible the integrals were evaluated analytically in the limit $|k_1 \eta_0| \gg 1$. Let us compute the squared norm of 
the modified initial-state shape function given by
\begin{eqnarray}
\left| F^\mathrm{modin}(k_1\eta_0,x_2,x_3) \right|^2 &=& \int_0^1 dx_2 \int_{1-x}^1 dx_3 \left[ 
F^\mathrm{modin}(k_1\eta_0,x_2,x_3) \right]^2 \, x_2^4 x_3^4 \nonumber \\
&=& \frac{\pi}{60} \, |k_1 \eta_0| + \frac{5}{4} \mathrm{log}|k_1 \eta_0| + 6.05 \, , \label{eq:modinshapenorm}
\end{eqnarray}
where the integral in $x_2, x_3$ space is over the (triangle) domain
$0 \leq x_2 \leq 1$, $1-x_2 \leq x_3 \leq 1$. From Eq.~(\ref{eq:modinshapenorm}) we can derive
some important conclusions about the detectability of this non-Gaussian signal. In an ideal situation
the data analysis would be performed using the theoretical template Eq.~(\ref{fmodin}) to directly infer on the non-Gaussian amplitude. As previously discussed this
is the product of the normalization $A$ times the norm of the shape function, $A |F^{modin}|$. Hence, the best one can
do by using an observational template perfectly aligned with the theoretical prediction is 
a leading enhancement factor of order $\sqrt{|k_1 \eta_0|}$. 
However, such an enhancement is lost when the data analysis is
performed using a local template, due to the integrated nature of the non-Gaussian analysis.
Evaluating the scalar product as defined by Eq.~(\ref{eq:scalarproduct})
between the initial-state modification template and the local one we find
\begin{eqnarray}
F^\mathrm{modin}(k_1\eta_0,x_2,x_3) \cdot F^\mathrm{local}(x_2,x_3) &=&\nonumber\\
\int_0^1 dx_2 \int_{1-x}^1 dx_3
F^\mathrm{modin}(k_1\eta_0,x_2,x_3) F^\mathrm{local}(x_2,x_3) \, x_2^4 x_3^4 
&=& 3 \mathrm{log} |k_1 \eta_0| + 18.96.  \label{eq:inlocalmodin}
\end{eqnarray}
We can already conclude that when using the local template to probe modified initial 
state non-Gaussianities the enhancement factor is further reduced to become
only logarithmic in $|k_1 \eta_0|$, instead of the $\sqrt{|k_1 \eta_0|}$ enhancement that can 
be achieved in the optimal case.
It is worth remarking that the constant parts in the results, 
for both the norm and the scalar product, depend on how some singular integrals, 
those independent of $k_1 \eta_0$, are being cut-off.
The singular integrals always blow up in the local 
(squeezed) limit, corresponding to one of the momenta being much smaller than the other two. 
Fortunately, a natural cut-off is given by the fact that only a finite range of modes contribute to the CMB.  Specifically, 
the ratio between the smallest and the largest observable scales on the CMB is roughly equal to $10^{-3}$, 
which is used to regularize the local type integrals. Throughout this paper 
we will always quote results using this cut-off if necessary. For the squared norm of the local shape, 
which clearly blows up in the squeezed limit, the need for this cut-off is most apparent. To be more explicit, 
using the $10^{-3}$ cut-off the result for the squared norm of the local shape equals $\left| F^\mathrm{local}(x_2,x_3) \right|^2 = 176.5$.
From the scalar product and the local and modified initial-state norms we can infer the cosine factor, which reads as
\begin{eqnarray}
\mathrm{Cos}\left[F^\mathrm{modin}, F^\mathrm{local}\right] 
&=& \frac{F^\mathrm{modin}(k_1\eta_0,x_2,x_3) \cdot F^\mathrm{local}(x_2,x_3)}
{\left| F^\mathrm{modin}(k_1\eta_0,x_2,x_3) \right| \left| F^\mathrm{local}(x_2,x_3) \right|} \nonumber \\
&=& 7.53 \cdot 10^{-2} \, \frac{(18.96 + 3 \log{|k_1 \eta_0|})}{\sqrt{6.05 + 
\frac{\pi}{60} |k_1 \eta_0| + \frac{5}{4} \log{|k_1 \eta_0|}}} \, . \label{eq:coslocmodin}
\end{eqnarray}
The fudge factor necessary to transform the limits on local type non-Gaussianities 
into constraints on modified initial-state non-Gaussianities is given by
\begin{eqnarray}
\Delta_F \left[F^\mathrm{modin}, F^\mathrm{local}\right] 
&=& \frac{F^\mathrm{modin}(k_1\eta_0,x_2,x_3) \cdot F^\mathrm{local}(x_2,x_3)}
{\left| F^\mathrm{local}(x_2,x_3) \right|^2} \nonumber \\
&=& 5.67 \cdot 10^{-3} \, (18.96 + 3 \log{|k_1 \eta_0|}) \, . \label{eq:fudgelocmodin}
\end{eqnarray}
As we already concluded from the scalar product alone, the fudge factor is logarithmically dependent on 
$k_1 \eta_0$. In addition, the coefficient is also relatively small, implying that over a realistic range
range of $|k_1 \eta_0|$ values, the fudge factor can essentially be considered constant.

\FIGURE{\includegraphics[width=130mm]{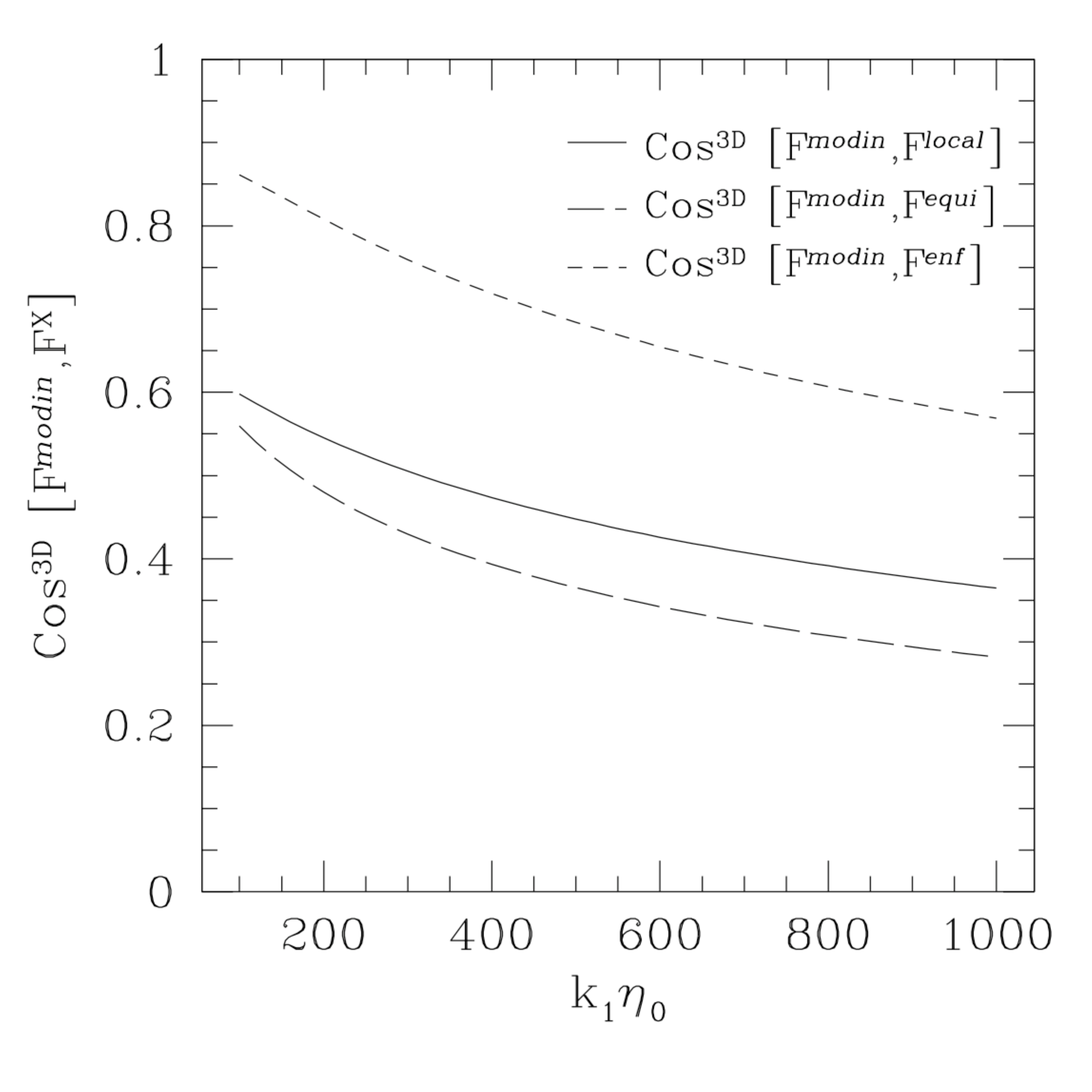}
\caption{\label{fig:cosplots} Cosines factors between the initial-state modification shape and the local (solid line),
equilateral (long dashed line) and the enfolded template proposal (short dashed line) as functions of $|k_1 \eta_0|$ in 3-D.}}

Similarly, we calculate the scalar product between the equilateral template and the 
modified initial-state distribution, which a priori can be expected to depend on $k_1 \eta_0$ as well. 
Surprisingly, the leading $k_1 \eta_0$ dependent terms cancel and the only contribution comes 
from a, cut-off independent, constant number for the scalar product
\begin{eqnarray}
F^\mathrm{modin}(k_1\eta_0,x_2,x_3) \cdot F^\mathrm{equil}(x_2,x_3) &=&\nonumber\\
\int_0^1 dx_2 \int_{1-x}^1 dx_3
F^\mathrm{modin}(k_1\eta_0,x_2,x_3) F^\mathrm{equil}(x_2,x_3) \, x_2^4 x_3^4
&=& 6.5 \, . \label{eq:inequilmodin}
\end{eqnarray}
We can therefore conclude that all enhancement is lost when using the equilateral template to 
probe modified initial-state non-Gaussianities. As for the local template this will imply a constant
fudge factor, even though the theoretical non-Gaussian distribution is linearly enhanced in enfolded triangles. 
The squared norm of the equilateral shape functions is also cut-off independent (i.e. finite), and
the numerical integration gives $\left| F^\mathrm{equil}(x_2,x_3) \right|^2 = 7.9$. Combining this with the
squared norm of the modified initial-state shape function this leads to the following expression 
for the normalization independent cosine 
\begin{eqnarray}
\mathrm{Cos}\left[F^\mathrm{modin}, F^\mathrm{equil}\right] 
&=& \frac{F^\mathrm{modin}(k_1\eta_0,x_2,x_3) \cdot F^\mathrm{equil}(x_2,x_3)}
{\left| F^\mathrm{modin}(k_1\eta_0,x_2,x_3) \right| \left| F^\mathrm{equil}(x_2,x_3) \right|} \nonumber \\
&=& \frac{2.31}{\sqrt{6.05 + \frac{\pi}{60} |k_1 \eta_0| + \frac{5}{4} 
\log{|k_1 \eta_0|}}} \, . \label{eq:cosequilmodin}
\end{eqnarray}
Even though the cosine is a function of $k_1 \eta_0$, while the scalar product 
is not, the fudge factor will also be independent and equals
\begin{eqnarray}
\Delta_F \left[F^\mathrm{modin}, F^\mathrm{equil}\right] 
&=& \frac{F^\mathrm{modin}(k_1\eta_0,x_2,x_3) \cdot F^\mathrm{equil}(x_2,x_3)}
{\left| F^\mathrm{equil}(x_2,x_3) \right|^2} \nonumber \\
&=& 0.82 \, . \label{eq:fudgequilmodin}
\end{eqnarray}
The constancy of the fudge factor explicitly confirms that all enhancement due to the large $|k_1 \eta_0|$ parameter
is lost. In Figure~\ref{fig:cosplots} and \ref{fig:fudgeplots} we plot the cosine and fudge 
factors between the initial-state modification and the local (solid line) and equilateral (long dashed line) templates 
as function of $k_1 \eta_0$. From the plot of the cosine factor we see that indeed the local and equilateral 
templates poorly overlap with the modified initial-state distribution as the $\cos[F^\mathrm{modin},F^{X}]<0.6$. 
We conclude that although the non-Gaussian amplitude of initial-state modifications is linearly enhanced 
in enfolded triangles, the measured local and equilateral templates are completely insensitive to this localized 
enhancement, thus spoiling any chance of obtaining a stringent bound on departures from the standard Bunch-Davies vacuum 
state.

\FIGURE{\includegraphics[width=130mm]{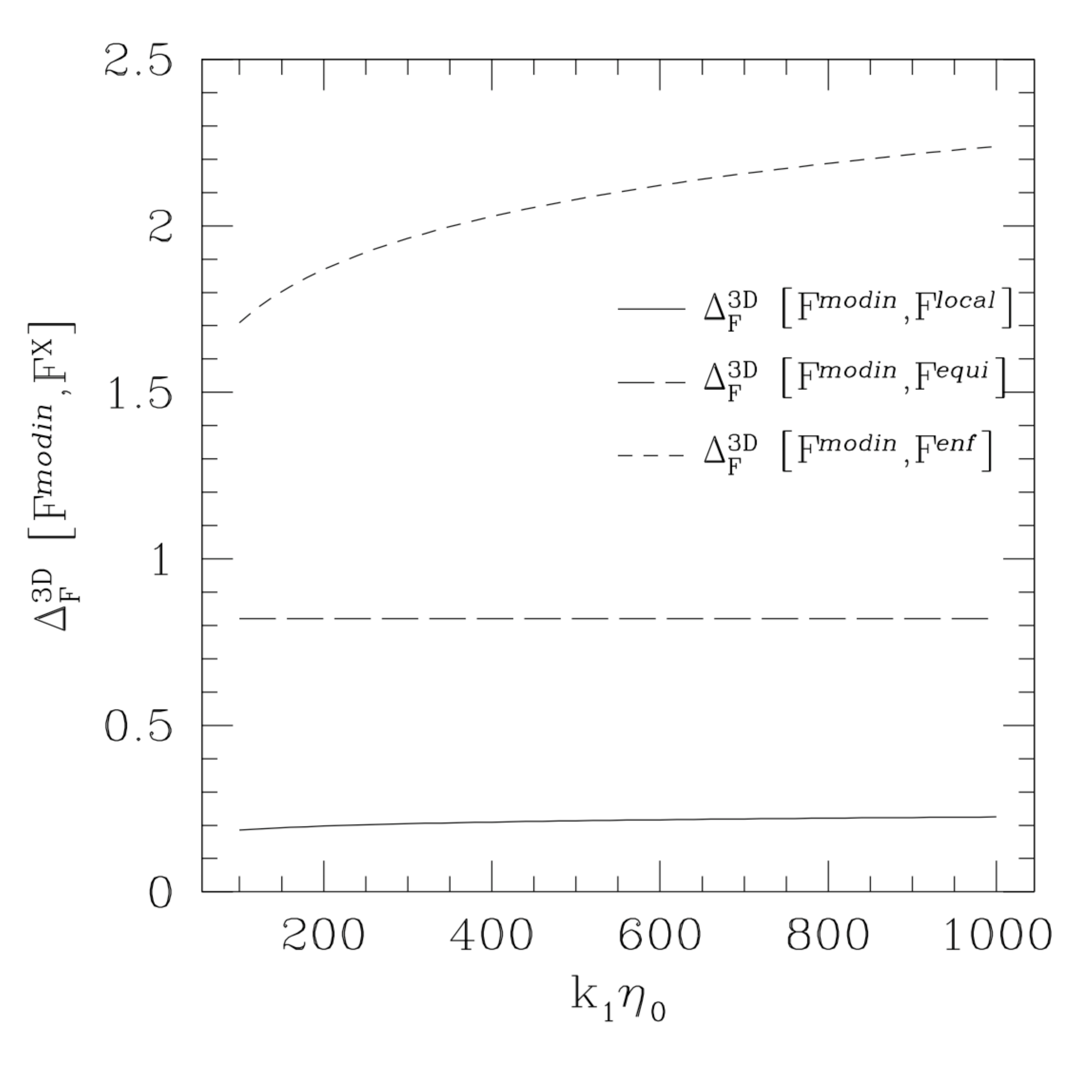}
\caption{\label{fig:fudgeplots} Fudge factors as functions of $|k_1 \eta_0|$ for the templates as 
in figure~$1$.}}

Consequently, probing standard slow-roll modified initial-state non-Gaussianities is impossible unless
a new template distribution is introduced which, unlike the local and equilateral templates, is sensitive to the localized 
enfolded enhancement. As pointed out in the previous discussion, using a perfect template will lead to a signal enhancement 
of $\sqrt{|k_1 \eta_0|}$. In Section~\ref{EnfTemp} we will describe a first proposal
for such an improved, factorized, enfolded template. In the next section we will focus on the combined effect of a specific 
higher derivative correction and an initial-state modification on the bispectrum.

\section{Adding higher derivative corrections}\label{HD}

As was shown in the previous section, the enhancement effect of an initial-state modification in the bispectrum, assuming standard slow-roll inflation, 
is impossible to probe using the currently available local or equilateral templates. What we would like to study is whether the same conclusion holds 
after adding higher derivative corrections, which according to \cite{Holman2007} could be even more sensitive to initial-state modifications. 
A priori, one might expect similar conclusions, that even though there is a strong enhancement effect in the enfolded triangle limit, 
its measure in the space of all triangles versus the local or equilateral template will again be too small to allow detection. 
We consider the addition of a dimension 8 higher derivative term to
the scalar field lagrangian of the following form:
\begin{eqnarray}
\Delta{\cal L}_{\mathrm{HD}}= \sqrt{-g} \frac{\lambda}{8M^4} \, \left( (\nabla \phi)^2 \right)^2 \, ,
\label{HDcorrection}
\end{eqnarray}
where the scale $M$ corresponds to the high energy cut-off scale and `natural' corrections would correspond to a coupling $\lambda \sim 1$.
This higher derivative correction is the same as the one discussed in
\cite{Holman2007} and was first studied in \cite{Creminelli2003}. We
provide a detailed derivation of the corresponding bispectrum in
Appendix B. As shown in \cite{Creminelli2003}, assuming the standard Bunch-Davies vacuum, it leads to
non-Gaussianities of the equilateral type with an amplitude $f_{\mathrm{NL}}^{\mathrm{equil}} \propto \left( \frac{M_{Pl}^2 H^2}{M^4} \right) \lambda \epsilon$ which, at best, can be of order $1$ (in order not to spoil the higher derivative expansion).
In the interaction Hamiltonian for the relevant perturbation variable $\zeta$ it leads to an additional term of the form
\begin{eqnarray}
\Delta H_{I}=-\frac{\lambda H}{2M^{4}}\int d^{3}xa(\eta)\left(\frac{\dot{\phi}}{H}\right)^{3}\zeta'\left(\zeta'^{2}-(\partial_{i}\zeta)^{2}\right).
\end{eqnarray}
As first shown in \cite{Holman2007}, and repeated here in the appendix, the associated bispectrum correction due to an initial-state modification
is a complicated function of the comoving momenta. Most importantly, compared to the result obtained for the standard slow-roll computation, after integrating over conformal time one now finds terms proportional to $\eta_{0}$ and $\eta_{0}^{2}$, in addition to contributions independent of $\eta_0$. 
The different powers of $\eta_{0}$ can, as before, be combined with one of the comoving momenta $k_1$ to give the large number $|k_1 \eta_0| = M/H \gg 1$. Consequently, the amplitude of the three-point function is expected to be dominated by contributions proportional to $\eta_{0}^{2}$. 
Collecting the leading contributions and neglecting terms that are not (locally) enhanced at $\eta_0^2$ order,  we obtain
\begin{eqnarray}
\langle\zeta_{k_{1}}\zeta_{k_{2}}\zeta_{k_{3}}\rangle_{\mathrm{nBD}}^{\mathrm{HD}} & \approx & (2\pi)^{3}\delta^{(3)}\left(\sum\vec{k}_{i}\right)\frac{\lambda}{M^{4}}\frac{1}{\prod(2k_{i}^{3})}\frac{H^{8}}{\dot{\phi}^{2}}\sum_{j}2\mathcal{R}e(\beta_{k_{j}}) \nonumber\\ 
&  & \times \left[  \left( \frac{1-\mathrm{cos}(\tilde{k}_{j}\eta_{0})}{\tilde{k}_j^2} 
- \eta_0 \, \frac{\mathrm{sin}(\tilde{k}_j\eta_0)}{\tilde{k}_j} \right) \, \mathcal{P}(k_j,k_{j+1},k_{j+2}) \right. \label{HDcontribution1} \\
&  & \left. + \eta_0^2 \left( \mathrm{cos}(\tilde{k}_j\eta_0) \, \mathcal{Q}(k_j,k_{j+1},k_{j+2}) \right) \right],
\label{HDcontribution2}
\end{eqnarray}
where
\begin{eqnarray*}
\mathcal{P}(k_{j},k_{j+1},k_{j+2}) & = & -4k_{j+1}k_{j+2}(k_{j+1}+k_{j+2})(k_{j+1}^{2}+k_{j+2}^{2}+k_{j+1}k_{j+2})\\
& & +2\times\prod_{i}k_{i}(k_{t}^{2}-4k_{j+1}k_{j+2}).\\
\mathcal{Q}(k_{j},k_{j+1},k_{j+2}) & = & \prod_{i}k_{i}(k_{t}^{2}-4k_{j+1}k_{j+2}).
\end{eqnarray*}
In the above expression $k_t=\sum_i k_i$ represents the sum of all (absolute values of) comoving momenta and $\tilde{k}_j$ is 
defined as before $\tilde{k}_j = k_t -2k_j$, i.e. the sum of comoving momenta with one of the signs reversed.
The terms in Eq.~(\ref{HDcontribution1}) are
enhanced to an order $\eta_0^2$ only in the collinear limit, while being suppressed for all other triangular configurations. 
In contrast the term in Eq.~(\ref{HDcontribution2}) is enhanced by $\eta_{0}^{2}$ over the full triangle domain and is therefore expected 
to be the dominant contribution. This was apparently not noticed in \cite{Holman2007}, maybe because the collinear
limit was assumed from the start. As a result the shape of the higher derivative bispectrum with a modified initial-state is 
not of the expected enfolded type. To be explicit let us rewrite the dominant overall enhanced 
contribution in terms of the two variables $x_2 \equiv \frac{k_2}{k_1}$ and $x_3 \equiv \frac{k_3}{k_1}$,  where we scaled out 
the usual $k_1^{-6}$ and absorbed the enhancement factor $(k_1 \eta_0)^2$ into the non-Gaussian amplitude $f_{\mathrm{NL}}$, 
producing the following shape function
\begin{eqnarray}
F^\mathrm{HD-dom}(k_1\eta_0,x_2,x_3)&=&\frac{1}{ x_2^2 x_3^2}\left\{ \cos(k_1 \eta_0(x_2 + x_3 -1)) \left[ (1+x_2 + x_3)^2 - 4 x_2 x_3 \right]
\right. \nonumber \\
& & + \cos(k_1 \eta_0(-x_2 + x_3 +1)) \left[ (1+x_2 + x_3)^2 - 4 x_3\right])  \nonumber \\
& &\left.  + \cos(k_1 \eta_0(x_2 - x_3 +1)) \left[ (1+x_2 + x_3)^2 - 4 x_2 \right]   \right\}.\label{eq:HDmodindom}
\end{eqnarray}
As before, scale-invariance of the bispectrum therefore requires the combination $|k_1 \eta_0| = M/H \gg 1$ to be $k_1$ independent.
Note that the cosines appearing in this shape function imply that the three-point function is constantly changing sign.
The norm of the full higher derivative modified initial-state distribution is well approximated using only the contribution described
by the shape function $F^\mathrm{HD-dom}$. We find, after averaging over the cosine, that $|F^\mathrm{HDmodin}|^2 \sim |F^\mathrm{HD-dom}|^2 \approx 23.3$\footnote{Remember that the overall $(k_1 \eta_0)^2$ enhancement factor was absorbed into the non-Gaussian amplitude 
$f_{\mathrm{NL}}$, explaining why it does not show up in the norm.}. Since it is the normalization independent combination 
$f_{\mathrm{NL}} \, |F|$ that is actually being measured, we conclude that a perfect template would be 
sensitive to the full $(k_1 \eta_0)^2$ enhancement factor. Because this number could be as large as $10^6$ it indicates that the higher 
derivative terms are extremely sensitive to initial-state modifications, potentially leading to strong constraints on departures from the 
standard Bunch-Davies vacuum state. This derives from the fact that the nonlinear higher derivative interaction 
plays an important role at sub-horizon scales. Sub-horizon particle occupation numbers as a consequence of the modified 
initial-state allow for the generation of a significant non-Gaussian signal due to the crucial presence of the higher derivative 
interactions at that stage. This is different from the standard slow-roll situation where the required (gravitational) nonlinearities 
are far less important at sub-horizon scales. 

Unfortunately though, due to the oscillating sign nature of the dominant contribution Eq.~(\ref{HDcontribution2}) the
currently available local and equilateral templates are extremely insensitive to this term, i.e. the scalar products between
$F^\mathrm{HD-dom}$ and the equilateral and local templates are suppressed because of cancellations inside the scalar product 
integral. A quick inspection of the scalar product integral reveals it could scale as $1/(k_1 \eta_0)$ times some oscillating function 
of $k_1 \eta_0$, which would reduce the overall $(k_1 \eta_0)^2$ level of enhancement by at least one power. 
It is for this reason that we have kept the locally enhanced terms of Eq.~(\ref{HDcontribution1}), since these could give rise to contributions 
in the scalar product of similar order in $k_1 \eta_0$.
The sine term in Eq.~(\ref{HDcontribution1}) is overall enhanced with one power of $k_1 \eta_0$ and, based on the results in the previous section,
the localized enhancement due  to the single $\tilde{k}_j$ in the denominator is expected to disappear after calculating the scalar product with the 
local or equilateral template, neglecting possible logarithmic terms. The cosine term is locally enhanced by a factor $(k_1 \eta_0)^2$ due to the 
squared $\tilde{k}_j$ dependence in the denominator. One of those enhancement factors is again expected to be lost after performing the scalar 
product integral with the local or equilateral templates. 

We anticipate a linear $k_1 \eta_0$ scaling at best (neglecting possible logarithmic terms) 
and we should keep track of all terms in Eq.~(\ref{HDcontribution1}) and Eq.~(\ref{HDcontribution2}) when evaluating the scalar product 
with the local or equilateral template. Unlike the previous section we were unable to perform the relevant integrals analytically and instead 
relied on a numerical approach, fitting the scalar product integral results for a large sample of $k_1 \eta_0$ values to estimate the $k_1 \eta_0$ dependence. The relevant shape function is identified in exactly the same way as the dominant contribution $F^\mathrm{HD-dom}$, except this 
time no overall factors of $k_1 \eta_0$ are absorbed into the definition of the non-Gaussian amplitude
\begin{eqnarray}
A &=& (2\pi)^4 \, \left( \frac{M_{Pl}^2 H^2}{M^4} \right) \, \frac{1}{2} \lambda \epsilon \, |\beta|  \,
\frac{\Delta_{\Phi}^{2}}{k_1^6}. \nonumber \label{eq:HDmodinamplitude}
\end{eqnarray}
As before we assume that the Bogoliubov parameter $\mathcal{R}e(\beta_{k_j}) \sim |\beta| $.
Having fixed the non-Gaussian amplitude and shape function ambiguities we find that the leading order behavior of 
the scalar product with the local template distribution is well approximated by
\begin{eqnarray}
F^{\mathrm{local}}\cdot F^{\mathrm{HDmodin}} \approx (k_1 \eta_0) \left(-72 + 10 \log|k_1\eta_0|  \right) \, , \label{eq:dotsinloc}
\end{eqnarray}
confirming the general expectation on the order of magnitude of the result. We should point out that the relative minus sign between the 
different terms in Eq. \eqref{HDcontribution1} is the source of the relative minus sign in the final result Eq.~\eqref{eq:dotsinloc}. 
Rather unfortunately, the different coefficients conspire in such a way that the scalar product has a minimum and then crosses through 
zero in the domain of interest $100 \leq  |k_1 \eta_0| \leq 1000$. 
This implies significantly smaller fudge factors, for a small range of $|k_1 \eta_0|$ values, than would 
be expected on the basis of scaling alone.  Another consequence of this, confirmed by the numerical results, is that the unavoidable oscillatory 
contributions, that we neglected when fitting the numerical data to produce Eq.~(\ref{eq:dotsinloc}), are bound to give rise to relatively 
large corrections in the $|k_1 \eta_0|$ domain of interest. The makes the full structure of the scalar product rather complicated. Although the 
general trend is nicely described by a linear plus logarithmic scaling with $|k_1 \eta_0|$, in the $|k_1 \eta_0|$ domain of interest the 
actual value of the scalar product fluctuates and can deviate from the expected order of magnitude for some values of $|k_1 \eta_0|$.
Dividing the scalar product by the norm of the local distribution (which was already computed in the previous section) one obtains the fudge factor
\begin{eqnarray}
\Delta_F \left[F^\mathrm{HDmodin}, F^\mathrm{local}\right] 
&=& \frac{F^\mathrm{HDmodin}(k_1\eta_0,x_2,x_3) \cdot F^\mathrm{local}(x_2,x_3)}
{\left| F^\mathrm{local}(x_2,x_3) \right|^2} \nonumber \\
&\approx& 5.7 \cdot 10^{-3} |k_1 \eta_0| \left(-72 +10 \log|k_1\eta_0| \right) \, . \label{eq:fudgelocal-HD}
\end{eqnarray}
We conclude that a $|k_1 \eta_0| \log |k_1 \eta_0|$ enhancement remains,  
which is almost one power of $k_1 \eta_0$ less as compared to the optimal scenario. For the fudge factor, as for the scalar product, the same cautionary remarks apply. 
The above result describes the average trend and the detailed numerical results show that fluctuations 
can have a significant effect on the actual value of the fudge factor in the $|k_1 \eta_0|$ domain of interest.
As we will see, the results for the 2d fudge factor with the local template exhibit a similar complicated behavior 
as a function of $k_1 \eta_0$, although the actual numbers for the fudge factor, due to the larger coefficients, 
are roughly one order of magnitude larger. For the equilateral template the final scaling result is the same, although somewhat
surprisingly, it is the dominant contribution in Eq.~(\ref{HDcontribution2}) that is solely responsible for the final result. 
As one can easily check, both terms in Eq.~(\ref{HDcontribution1}) are in fact maximizing exactly at the line $x_2+x_3-1=0$, 
whereas the equilateral template is exactly vanishing at the line $x_2 + x_3-1=0$. The result is a suppressed contribution 
to the scalar product which is negligible in the limit of large $|k_1 \eta_0|$ as compared to the other contribution. 
Incidentally this observable enhancement is the same as reported in \cite{Holman2007}, 
but the underlying reason is very different. 
It is a consequence of using a template that is far from optimal and it should be possible to achieve significantly higher sensitivity 
by constructing a more suitable template to analyze the data. In particular, the non-Gaussian signal 
described by Eq.~(\ref{HDcontribution1}) and Eq.~(\ref{HDcontribution2}) is not of the enfolded type and 
has strong oscillatory features, which might allow for a clear distinction from other non-Gaussian sources. 

The generic appearance of at least a single factor of $|k_1 \eta_0| = M/H$ in the fudge factor with respect to the local (or equilateral) 
template implies an enhancement possibly as large as $10^3$, ignoring the fluctuations of the fudge factor as a function of 
$k_1 \eta_0$.  At the start of this section we mentioned that in the standard 
Bunch-Davies vacuum the higher derivative term would give rise to a maximal $f_{\mathrm{NL}}^{\mathrm{equil}}$ of order $1$. 
Compared to the original higher derivative non-Gaussian amplitude, the modified initial-state amplitude Eq.~(\ref{eq:HDmodinamplitude})
introduces an additional suppression with the Bogoliubov parameter. On the other hand the fudge factor introduces a linear 
$|k_1 \eta_0| \equiv M/H$ factor enhancing the original higher derivative non-Gaussian amplitude 
by $\beta \, (M/H)$ when probed with the local or equilateral template. The CMB two-point power spectrum constrains the 
Bogoliubov parameter already at the $10^{-2}$ level, so at best this would allow for a local or equilateral non-Gaussian 
amplitude of order $10$ due to initial-state modifications, assuming $M/H \sim 10^3$. This might be detectable in the future, although
there are many other sources for a local or equilateral non-Gaussian signal at that level. In section \ref{CMB} we will confirm the 
same level of enhancement by computing the projected 2d fudge factor and use the most recent WMAP constraints on 
local type non-Gaussianities to derive an order of magnitude constraint on the Bogoliubov parameter.

\section{An enfolded template proposal}\label{EnfTemp}

%{\bf Perhaps move this to an appendix, to downplay its relevance.}

\FIGURE{
\includegraphics[width=140mm]{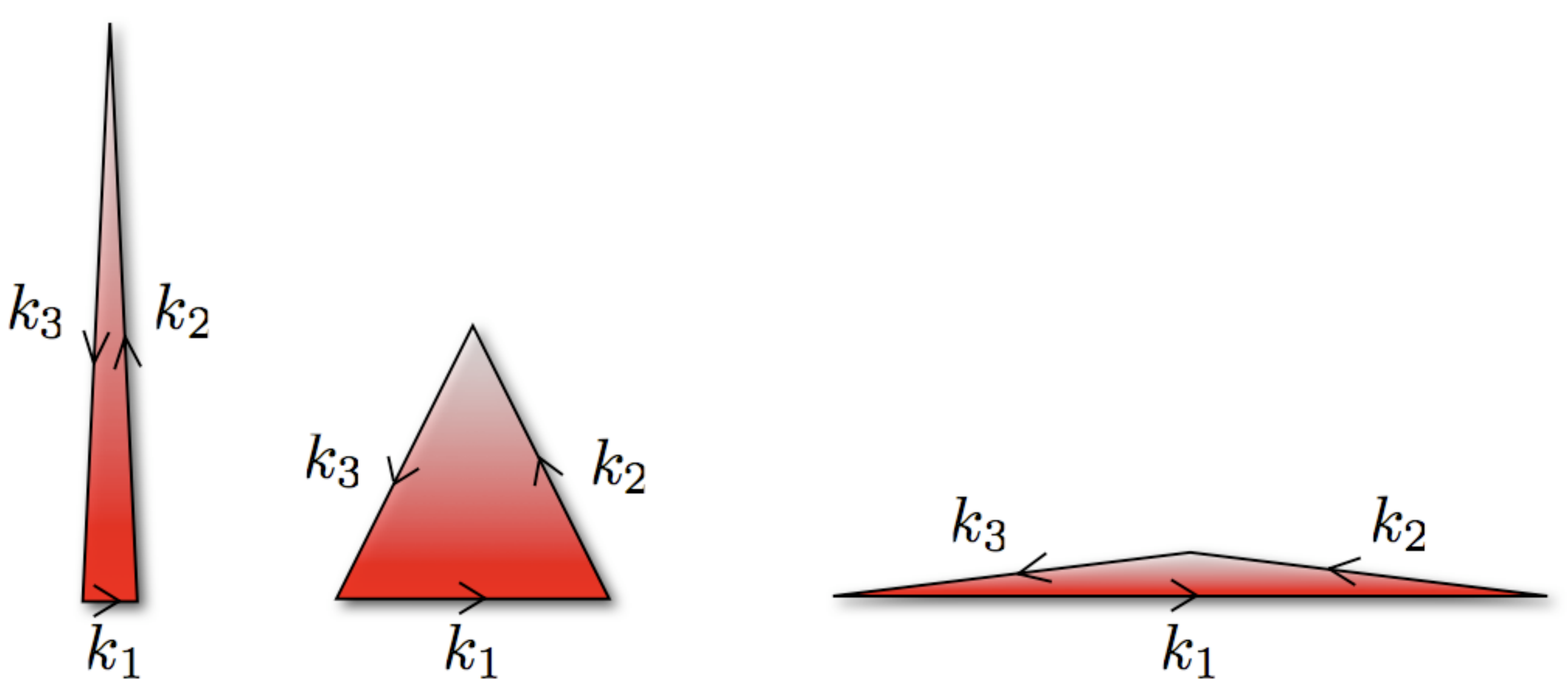}
\caption{\label{fig:triangles}From left to right: a squeezed, equilateral and squashed triangle.}
}

In the absence of higher derivative corrections we have shown that a non-Gaussian signal 
due to a modified initial-state, which maximizes in collinear triangles, cannot be probed using
the available local and equilateral templates. Both templates are not sensitive enough to the localized enfolded enhancement
to give rise to a significant (preferably power law) dependence of the fudge factor for large 
$|k_1 \eta_0|=M/H$. Instead the local and equilateral fudge factors are at best logarithmically
dependent on $|k_1 \eta_0|$. To see if one can improve on that situation one would like to introduce 
a more suitable template, one better aligned with the theoretical prediction of modified initial-state 
non-Gaussianities.

The distinguishing feature of such a template should be that it maximizes in the enfolded (or squashed) 
triangle limit, as opposed to the local and equilateral templates which maximize in squeezed and equilateral 
triangles respectively. Whereas the squeezed triangle is obtained by taking one of the
comoving momenta to zero $k_{i}\rightarrow0$,  and equilateral corresponds to all momenta equal, 
enfolded triangles imply two collinear momenta, and therefore $k_i=k_j+k_m$ with $i\neq j\neq m$ and
$k_j$ and $k_m$ representing the two collinear momenta.  Clearly the squeezed, equilateral and squashed triangle
limits exhaust all possibilities, which are shown in figure \ref{fig:triangles}, and nicely correspond to the 
three different classes of theoretical non-Gaussian predictions: local, equilateral and enfolded. So besides its potential
theoretical relevance, also from the point of view of completeness it might be worthwhile to develop a third factorized 
template shape that would maximize in the enfolded triangle limit. This would introduce a third non-Gaussian observable  
$f_{\mathrm{NL}}^{\mathrm{enf}}$ measuring the enfolded amplitude.

The reason why one cannot directly compare theoretical predictions to the CMB data and needs 
especially designed templates lies in the computational complexity of reconstructing the non-Gaussian amplitude
from the two-dimensional CMB temperature data. The projection of a $3$-point correlator to a $3$-point function in 
spherical harmonic space involves the Wigner $3j$ symbol (to construct the 
angular averaged bispectrum) and a complicated integral over transfer and Bessel functions. 
This is computationally very challenging, scaling as $N^{5/2}$; $N^{1/2}$ for every multipole $l$ and $N$ 
for the averaging over $m$, where $N$ equals the total number of pixels in the CMB map. 
In the last few years different suggestions have been made to accomplish a reduction
of computational time \cite{Komatsu2003a, Komatsu2003, Babich2004a, Liguori2005, Kogo2006, 
Creminelli2006, Yadav2007a, Liguori2007, Yadav2007b}. A significant reduction 
in the number of calculations can be achieved if the three-point function is factorizable in its
momentum dependence, i.e. $F(k_{1},k_{2},k_{3})\rightarrow f_{a}(k_{1})f_{b}(k_{2})f_{c}(k_{3})$,
leading to a reduction from $N^{5/2}$ to $N^{3/2}$. As it turns out, local type non-Gaussianities are 
indeed described by a factorized shape function $F(k_1,k_2,k_3)$, whereas the theoretical predictions 
for non-Gaussianities of the equilateral and enfolded type are not factorizable. This makes the direct comparison of equilateral and enfolded 
type non-Gaussianities to the two-dimensional CMB data extremely difficult for
the time being, although recently some progress has been made to allow for a more 
direct comparison of arbitrary signals \cite{Fergusson2006, Fergusson2008, Smith2006}. 
Up to now one instead relies on factorized template approximations to the theoretical signals. Not so long ago an equilateral template has 
been successfully identified and compared to the data \cite{Komatsu2008, Creminelli2005, CSZT2007, Creminelli2006}, 
but an enfolded observational template has not yet been constructed. Below we will construct 
a first proposal for a factorized enfolded template and analyze how much better it will be able
to constrain the modified initial-state non-Gaussianities discussed in section 3. 

\FIGURE{
\includegraphics[width=140mm]{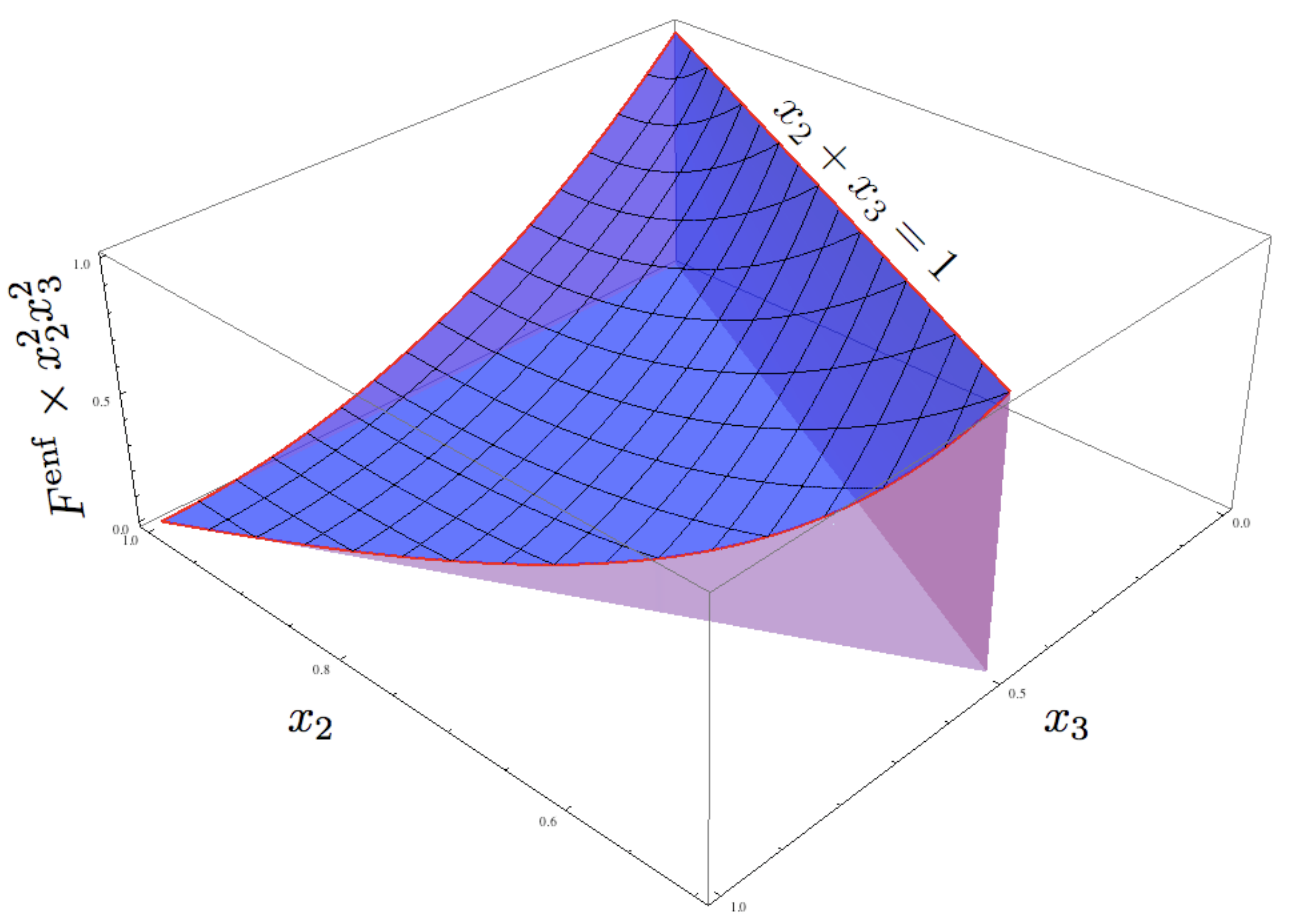}
\caption{\label{fig:enfoldedtemplate} The enfolded template shape
$F(x_2,x_3)\times x_2^2 x_3^2$.}
}

Looking at Eq.~\eqref{fmodin} it is clear that the three-point correlation function due to initial-state modifications
is not factorizable. As explained this non-Gaussian shape function is the result of adding a minus sign to 
one of the comoving momenta to first order in the Bogoliubov parameter $\beta_{k}$  \cite{single_field_ng_2, Holman2007}. 
For instance a term behaving as $1/(k_{1}+k_{2}+k_{3})$ would change to $1/(-k_{1}+k_{2}+k_{3})$ plus permutations.
This suggests an approach where one starts with the factorized equilateral template eq.\eqref{eq:equilshape}
and just replaces $k_i\rightarrow-k_i$, symmetrizing over all the indices. Applying this idea produces the shape function
$F(k_1,k_2,k_3)= - F_{\mathrm{equil}}(k_1,k_2,k_3)$, which does not yet resemble the desired enfolded distribution nor does it add additional information, i.e. it is simply the 
equilateral shape multiplied by a minus sign. Fortunately 
though it requires only a small modification to come up with a factorized shape function that seems to be a reasonable candidate for 
an enfolded template. Starting from the equilateral shape function, replacing $k \rightarrow -k$, introducing $x_2=k_2/k_1$ and $x_3=k_3/k_1$ 
and plotting the obtained distribution $F(x_2,x_3)$ times the appropriate measure factor\footnote{This
is the relevant quantity because of the measure in the scalar product of eq. \eqref{eq:scalarproduct}} $x_2^2 \, x_3^2$,
it becomes apparent that a term proportional to $1/k_{1}^{2}k_{2}^{2}k_{3}^{2}$ acts as a kind of constant `normalization' of the 
template. Additional $1/k_{1}^{2}k_{2}^{2}k_{3}^{2}$ terms therefore simply {\it lift} or {\it lower} the whole
graph. By adjusting the number of such terms, so $F^{\mathrm{enf}}=-F^{\mathrm{equil}}+c/k_1^2k_2^2k_3^2$, 
it is possible to lift the obtained shape in such a way that it resembles an enfolded type distribution, maximizing 
on the line $k_2+k_3\simeq k_1$, corresponding to enfolded triangles. We find that the best choice requires 
adding only one such term to the $-F_{\mathrm{equil}}$ distribution, i.e., $c=1$ (see Appendix A). 
Consequently our proposal for the factorized enfolded template, as a function of $x_2$, $x_3$, becomes
\begin{eqnarray}
F^{\mathrm{enf}}(x_{2},x_{3}) & = & 6 \left[\frac{1}{x_{2}^{3}}+2\;\mathrm{perm}+\frac{3}{x_{2}^{2}x_{3}^{2}}
-\left(\frac{1}{x_{2}^{2}x_{3}^{3}}+5\;\mathrm{perm}\right)\right].
\label{eq:enfoldedtemplate}
\end{eqnarray}
We have plotted the template shape function in figure \ref{fig:enfoldedtemplate}. 
In appendix A we explain in what sense $c=1$ corresponds to the optimal choice and the 
details are presented on how to translate this template into an observable using the `fast best estimator' approach 
developed in \cite{Komatsu2003a, Komatsu2003, Creminelli2006, Yadav2007a, Yadav2007b}.

We should determine how well the proposed enfolded template overlaps with the theoretical 
modified initial-state three-point function of Eq.~\eqref{fmodin}. To quantify this we will perform 
the same analysis as in section 3, calculating the scalar product, cosine and fudge factor, 
now using the enfolded template. The squared norm of the modified initial-state shape function is given by Eq.~(\ref{eq:modinshapenorm}) and a numerical integration gives $|F^{\mathrm{enf}}(x_{2},x_{3})|^2 =4.34$. After computing the scalar product, 
this leads to the following expression for the cosine as a function of $|k_1 \eta_0|$
\begin{eqnarray}
\mathrm{Cos}\left[F^\mathrm{modin}, F^\mathrm{enf}\right] 
&=& \frac{F^\mathrm{modin}(k_1\eta_0,x_2,x_3) \cdot F^\mathrm{enf}(x_2,x_3)}
{\left| F^\mathrm{modin}(k_1\eta_0,x_2,x_3) \right| \left| F^\mathrm{enf}(x_2,x_3) \right|} \nonumber \\
&=& 4.80 \cdot 10^{-1} \, \frac{(2.80 + \log{|k_1 \eta_0|})}{\sqrt{6.05 + \frac{\pi}{60} |k_1 \eta_0| + \frac
{5}{4} \log{|k_1 \eta_0|}}} \, . \label{eq:cosenfmodin}
\end{eqnarray}
We have plotted this function in figure \ref{fig:cosplots}, including the cosine functions between the 
modified initial-state shape distribution and the local and equilateral templates. As should be clear from 
the plot the cosine between the enfolded template and the theoretical distribution is 
closer to one, but there is certainly room for improvement. As the parameter $|k_1 \eta_0|$ grows, the 
enfolded template will depart more from the theoretically predicted modified initial-state shape. 
Nevertheless, the enfolded template has significantly higher overlap with the theoretical distribution than
the local or equilateral templates, at least in the comoving (3-D) momentum space. Since the cosine is the relevant quantity that determines the 
relative improvement, comparing to the plots (see also table 1) for the local and equilateral cosines one concludes 
that a rough $35-45$ percent level of improvement should theoretically be achievable using the enfolded template.
This is certainly not enough to derive interesting constraints for the theoretically predicted 
modified initial-state non-Gaussianities, as can be seen more directly by turning our attention to the fudge factor,
which explicitly identifies the level of $k_1 \eta_0$ enhancement. From the expression of the cosine it is straightforward 
to read off the fudge factor 
\begin{equation}
\Delta_F \left[ F^{\mathrm{modin}}, F^{\mathrm{enf}} \right] = 0.65 + 0.23 \log{|k_1 \eta_0|}
\end{equation}
which disappointingly implies that the enhancement, or the sensitivity, that would be achieved using the 
proposed enfolded template is still only logarithmic in $|k_1 \eta_0|$, far removed from the 
maximally attainable level of $\sqrt{|k_1 \eta_0|}$ enhancement. Comparing to the local and equilateral
fudge factors the coefficients are bigger which means the enfolded fudge factor will be larger
and changes considerably over the natural range of $|k_1 \eta_0|$ ($10^2 - 10^3$). 
This is however still far removed from the (power law) level of enhancement that one would need to derive interesting 
constraints on the Bogoliubov parameter from the bispectrum data. 

\TABLE{
\begin{tabular}{|c|c|c|c|c|}
     \hline 

Shape $F_Y$ & 3-D Cos ($F_X=F_\mathrm{loc}$) & 3-D Cos ($F_X=F_\mathrm{eq.}$) & 3-D Cos ($F_X=F_\mathrm{enf}$)\\
     \hline 
     Local  & 1 & 0.41 & 0.68 \\
     Equil. & 0.41 & 1 & 0.49  \\
     Enf. & 0.68 & 0.49 & 1\\
   \hline
    HD  & 0.45 & 0.99 & 0.59 \\
    Mod & 0.6 -- 0.3 & 0.6 -- 0.4  & 0.9 -- 0.6  \\
     \hline
\end{tabular}
\caption{\label{cos-results} \small The 3d Cosine}
}

\TABLE{
\begin{tabular}{|c|c|c|c|c|}
     \hline 
     Shape $F_Y$ & 3-D Fudge ($F_X=F_\mathrm{loc}$) & 3-D Fudge ($F_X=F_\mathrm{eq.}$) & 3-D Fudge ($F_X=F_\mathrm{enf}$)\\
     \hline 
     Local  & 1 & 1.94 & 4.29 \\
     Equil. & 0.09 & 1 & 0.66  \\
     Enf. & 0.11 & 0.36 & 1\\
     \hline
    HD  & 0.10 & 1.07 & 0.86 \\
    Mod & $ 0.11 + 0.017 \log{|k_1 \eta_0|}$  & 0.82 & $0.65+ 0.23 \log{|k_1 \eta_0|}$  \\
     \hline
\end{tabular}
\caption{\label{ff-results} \small The 3d Fudge Factors. Note that HD are the higher derivative contributions from \cite{Babich2004a}. We added these for 
completeness and to show consistency with the results in \cite{Babich2004a}.}
}

In tables \ref{cos-results} and \ref{ff-results} we have collected the cosine and fudge factor results for the different templates with respect to 
each other and some important theoretical predictions (equilateral higher derivative and enfolded modified initial-state). Note that the 
modified initial-state entries typically depend on the (large) parameter $k_1 \eta_0$, which for the cosines has been denoted by the range
of possible values, whereas for the fudge factors we have chosen to explicitly write the function. Independent from the original 
theoretical motivation, one could argue that the enfolded template nicely completes a general analysis of non-Gaussian signals. 
Table \ref{cos-results} then shows how much complementary information each template would provide. 
Compare this to a decomposition of a general vector into a set of basis vectors. Ideally, one would prefer to  
come up with a set of orthogonal basis shapes. Instead, the local, equilateral and enfolded template are far from orthogonal, but each does 
provide complimentary information that can be precisely quantified in terms of the different cosine values\footnote{Note that since we are 
dealing with a function space, obviously a complete decomposition would formally require an infinite set of basis functions on the relevant 
triangle domain.} listed in the table. Decomposing a general three-point signal in these template shapes might therefore still be useful, 
even though the enfolded template by itself is unable to probe modified initial-state non-Gaussianities. 

\section{Two-dimensional bispectrum results}\label{CMB}
Differently from the 3-D case the CMB temperature
anisotropies are a 2-D projection of the linearly evolved primordial
curvature perturbation field, hence the 2-D bispectrum is the result 
of the convolution of the shape function with the photon transfer 
function projected on the sky. In the following we give a brief 
review of the basic formalism, then we will discuss the results of 
the CMB bispectrum computation.

Let us consider the standard
spherical harmonic decomposition of the CMB temperature
fluctuation along the direction $\hat{n}$ of the sky,
\begin{equation}
\frac{\Delta T}{T}(\hat{n})=\sum_{l,m} a_l^m Y_l^m(\hat{n}),
\end{equation}
the multipole coefficients $a_l^m$ contain all statistical information 
about the temperature anisotropy field, and are the starting point to 
construct the various correlator functions. 
The angular bispectrum in multipole space is given by
\begin{equation}
B_{l_1 l_2 l_3}^{m_1 m_2 m_3} \equiv <a_{l_1}^{m_1}a_{l_2}^{m_2}a_{l_3}^{m_3}>,
\end{equation}
and assuming rotational invariance, the angle-averaged bispectrum
reads as 
\begin{equation}
B_{l_1 l_2 l_3}=\sum_{m_1,m_2,m_3}\left(\begin{array}{ccc}
l_1 & l_2 & l_3 \\
m_1 & m_2 & m_3 \end{array}
\right)<a_{l_1}^{m_1}a_{l_2}^{m_2}a_{l_3}^{m_3}>.\label{bispgen}
\end{equation} 
Substituting the expression of the multipole coefficients in terms of the photon
transfer function and the primordial curvature perturbation, 
Eq.~(\ref{bispgen}) becomes 
\begin{equation}
B_{l_1 l_2 l_3}=\sqrt{\frac{(2 l_1+1)(2
 l_2+1)(2 l_3+1)}{4\pi}}\left(\begin{array}{ccc}
l_1 & l_2 & l_3 \\
0 & 0 & 0 \end{array}\right)b_{l_1 l_2 l_3},\label{eq:Blll}
\end{equation}
with $b_{l_1 l_2 l_3}$ the reduced bispectrum given by,
\begin{eqnarray}
b_{l_1 l_2 l_3}=\left(\frac{2}{\pi}\right)^3 \int dx dk_1 dk_2 dk_3 (x
k_1 k_2 k_3)^2 j_{l_1}(k_1 x) j_{l_2}(k_2 x) j_{l_3}(k_3 x) \nonumber \\
\times F(k_1,k_2,k_3) \Delta_{l_1}(k_1) \Delta_{l_2}(k_2)
\Delta_{l_3}(k_3),\label{redbi}
\end{eqnarray}
where $\Delta_{l_i}(k_i)$ is the photon transfer function and
$j_{l_i}(k_i x)$ is the Bessel function (for a detailed derivation
see \cite{Komatsu2001a}).
As discussed in \cite{Babich2004a}, the evaluation of the reduced
bispectrum is computationally
challenging, on the other hand in the flat-sky approximation
the computation is simplified, since 
for example the integral over the Bessel function does not explicitly
appear. In such a case one has:
\begin{eqnarray}
b_{l_1 l_2 l_3}=\frac{(\tau_0-\tau_R)^2}{(2\pi)^2}\int dk_1^z dk_2^z
dk_3^z \delta(k_1^z+k_2^z+k_3^z) F(k_1',k_2',k_3')
\tilde{\Delta}_{l_1}(k_1^z) \tilde{\Delta}_{l_1}(k_2^z)
\tilde{\Delta}_{l_1}(k_3^z),\nonumber \\ \label{redbiflat}
\end{eqnarray}  
where $\tau_0$ and $\tau_R$ are the conformal time today and at
decoupling respectively, $k_i^z$ is the component of the wave-vector
orthogonal to the plane tangent to the last scattering surface, and $k'=\sqrt{(k^z)^2+l^2/(\tau_0-\tau_R)^2}$;
the photon transfer function along the orthogonal direction is given by
\begin{equation}
\tilde{\Delta}_{l}(k^z)=\int \frac{d\tau}{(\tau_0-\tau)^2}S(\sqrt{(k^z)^2+l^2/(\tau_0-\tau)^2})e^{i k^z \tau},
\end{equation}
where $S(...)$ is the CMB source function.
The presence of the delta-function ensures that the projected
modes form closed triangles. Hereafter we assume a vanilla LCDM cosmology
with model parameter values corresponding to the WMAP 5-years best-fit model.
All cosmologically relevant quantities such as the source functions
have been computed with the publicly available CMBFAST code 
\cite{Zaldarriaga}. We then have evaluated the reduced bispectrum 
for different shapes using Eq.~(\ref{redbiflat}).

Following \cite{Babich2004a} we introduce a scalar product
\begin{equation}
B_X \cdot B_Y = \sum_{l_1,l_2,l_3} \frac{B_{l_1 l_2 l_3}^X B_{l_1 l_2 l_3}^Y}{f_{l_1 l_2 l_3}C_{l_1}C_{l_2}C_{l_3}}, \label{eq:BB}
\end{equation}
where $f_{l_1 l_2 l_3}$ is a combinatorial factor which is $1$ if all three multipoles are different, $2$ if two of them are equal
and $6$ if all of them are equal; $C_l$ is the angular CMB power spectrum which includes the experimental noise evaluated in the 
Gaussian approximation assuming WMAP experimental characteristics. The cosine reads as
\begin{equation} 
\mathrm{Cos}^{2D}[B_X,B_Y]=\frac{B_X \cdot B_Y}{\sqrt{B_X \cdot B_X}\sqrt{B_Y \cdot B_Y}},
\end{equation}
and the fudge factor as
\begin{equation} 
\Delta_F^{2D}[B_X,B_Y]=\frac{B_X \cdot B_{\mathrm{local}}}{B_{\mathrm{local}} \cdot B_{\mathrm{local}}}.
\end{equation}

In principle we may expect that the non-Gaussian signal given by different triangular shapes on the CMB 
differs from that obtained in 3-D, since triangles of different shapes in 3-D can be projected into
the same 2-D configuration. If this is the case then the values of the cosine factors should be shifted upwards.
Evaluating the cosine and fudge factors between local and equilateral
shapes we find ${\rm Cos}^{\rm 2D}[F^{\mathrm{local}},F^{\mathrm{equil}}]=0.62$
and $\Delta_F^{\rm 2D}[F^{\mathrm{local}},F^{\mathrm{equil}}]=0.13$ respectively, which is consistent with the results presented in \cite{Babich2004a}. The cosine and fudge factors between the initial-state modification shape function and the local, equilateral and enfolded
templates are shown in figure~\ref{fig:cos2d} and \ref{fig:fudge2d} as a function of the $k_1\eta_0$ parameter respectively.

\FIGURE{
\includegraphics[width=140mm]{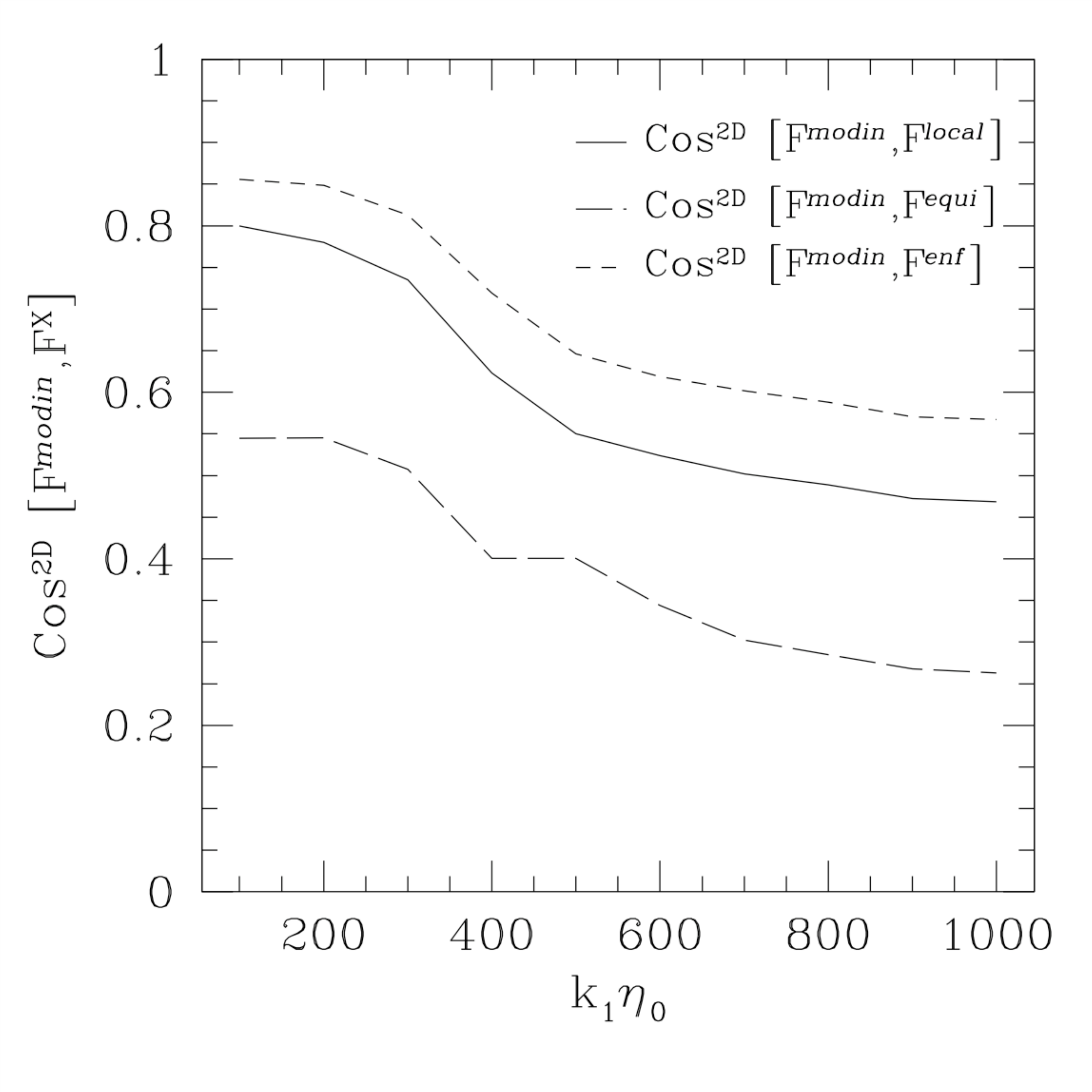}
\caption{\label{fig:cos2d}2-D cosine factors between the initial 
state modification template and the local (solid line), equilateral (long dashed line)
and enfolded (short dashed line).}
}

We may notice a trend similar to that inferred from the 3-D evaluation. In particular, the cosine decreases 
as a function of $k_1\eta_0$ for all three templates, whereas the fudge factors are 
constant for the local and equilateral case, and increasing for the enfolded template. 
The enfolded template has the largest overlap with the initial
state modification shape, although not significantly better then the local one. 
Overall the cosine values are slightly larger than what we have found in the 3-D calculation. 
This is because different triangular configuration in 3-D can be degenerate in the 2-D,
hence the projection tends to systematically increase the overlapping
between different templates.

\FIGURE{
\includegraphics[width=140mm]{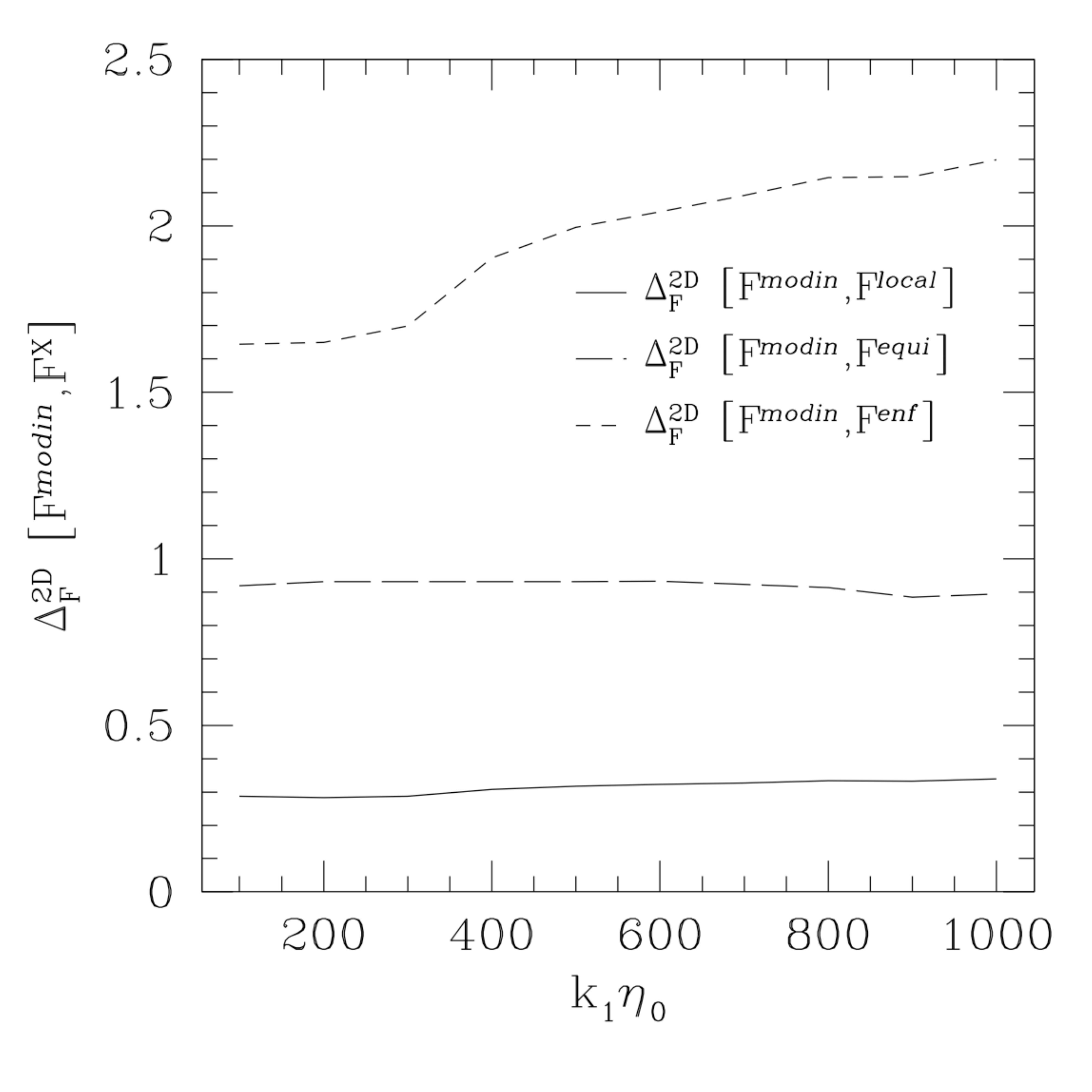}
\caption{\label{fig:fudge2d}2-D fudge factors between the initial 
state modification template and the local (solid line), equilateral (long dashed line)
and enfolded (short dashed line).}
}

We have also computed the reduced CMB bispectrum for the initial-state
modification in the presence of higher order derivative terms as given
by the shape function Eq.~(\ref{HDcontribution1}) and Eq.~(\ref{HDcontribution2}). We confirm
the enhancement effect discussed in section~\ref{HD}, as an example
calculating the fudge factor with the local template for 
$k_1\eta_0=100$ we find $|\Delta^{\rm 2D}_F|\approx 100$, whereas for 
$k_1\eta_0=10^3$ we obtain $|\Delta^{\rm 2D}_F|\approx
6000$. This clearly shows that the non-Gaussianity induced by initial
state modifications is enhanced by the presence of higher derivative
terms leading to potentially large detectable signals.
The specific functional dependence of the fudge factor on $k_1\eta_0$ 
in the range of interests ($10^2-10^3$) is far from trivial due to 
the interplay of the different terms in
Eq.~(\ref{HDcontribution1}) and Eq.~(\ref{HDcontribution2}). 
These contain oscillatory factors 
that leads to a modulated oscillations of the fudge
factor dependence on $k_1\eta_0$.  Besides, the same term causes an
oscillatory dependence of the reduced bispectrum as function of 
the multipoles. These oscillations are responsible for cancellations 
in the evaluation of the cosine factor, hence leading
to a very small overlap with the other (non-oscillatory) templates.
For example evaluating the cosine factor with the local shape for 
$k_1\eta_0$ in the range $10^2-10^3$ we find $|{\rm Cos}^{\rm 2D}_F|\approx 0.01$, 
implying that current observational templates are not apt to
detect such a non-Gaussian signal. 
The determination of an optimal observational
template that can account for the feature produced by this type of 
non-Gaussianity is therefore necessary and we leave this search to future
work. Nevertheless using the above estimates of the fudge factor, we can
use current constraints on local non-Gaussianity to infer limits on $f_{\rm NL}^{\mathrm{HDmodin}}$ 
and thus on the Bogoliubov parameter. As an example the prediction for a local non-Gaussian 
contribution due to a modified initial-state in the presence of a higher derivative operator reads 
\begin{equation}
|\Delta f_{\mathrm{NL}}^{\mathrm{local}}| = \frac{5}{6} \epsilon |\beta| \left( \frac{M_{Pl}^2 H^2}{M^4} \right) \,  |\Delta^{\rm 2D}_F| \, ,
\end{equation}
where we assumed that the coefficient $\lambda$ in Eq.~(\ref{eq:HDmodinamplitude}) equals one.
Using the observed amplitude of the power spectrum, the slow-roll parameter $\epsilon$ can be replaced with
$\frac{10^{10}}{8\pi^2} \, \frac{H^2}{M_{Pl}^2}$. Assuming $M/H \sim 10^3$ this gives rise to the following constraint on 
the Bogoliubov parameter, using the latest WMAP 5-years upper limit on $f_{\rm NL}^{\rm local}$ and the result for 
the fudge factor $|\Delta^{\rm 2D}_F| \approx 6000$
\begin{equation}
4 \cdot 10^3 \,  |\beta| < 111 \, .
\label{betaconstraint}
\end{equation}
Surprisingly, this corresponds to a relatively strong bound on the Bogoliubov parameter $|\beta| < 3 \cdot 10^{-2}$.
Note that in most proposals one expects the Bogoliubov parameter $|\beta|$ to be a function of $H/M$. For instance
in the New Physics Hypersurface scenario $|\beta|$ is predicted to be linear in $H/M$ \cite{Danielsson2002, Easther2002} . 
Using the results above for $H/M \sim 10^{-3}$ this predicts at best an order $10$ contribution to the local non-Gaussian signal, 
which could increase to an order $10^2$ contribution for $H/M \sim 10^{-2}$.  We should stress that an ideal template could 
improve the limits on the Bogoliubov parameter by another factor of $M/H$.

\section{Conclusion}\label{Conclusion}

We have analyzed inflationary three-point correlators as a result of a small departure from the standard Bunch-Davies vacuum. 
In the simplest scenario where we avoided higher derivative interactions we confirmed that the initial-state modification causes 
the three-point correlator to maximize in the collinear or enfolded triangle limit, corresponding to a uniquely different shape as  
compared to local and equilateral non-Gaussian signals. Since the maximal signal scales linearly with the 
cut-off scale $M$, the non-Gaussian amplitude in enfolded triangles can be quite large, perhaps allowing for detection or 
providing interesting constraints on departures from the Bunch-Davies vacuum. However, by computing the scalar 
products, and consequently the cosine and fudge factors, between the theoretical prediction for the three-point function 
and the existing local and equilateral observational templates, we concluded that essentially all enhancement 
is lost due to the inefficiency of the available observational templates combined with the localized nature of the 
enhancement. Although \cite{Holman2007} reached a similar conclusion, their argument was very different, relying on the 
projection to the two-dimensional CMB sphere. Instead, we have shown that the currently available method of comparing theoretical 
three-point functions to CMB bispectra, involving observational templates and the necessary integration over all triangles, already 
removes most sensitivity to localized enhancements in enfolded triangles, even before projecting to the two-dimensional CMB sphere.

The situation can in principle be improved by constructing a suitable enfolded template (better) adapted to the theoretical 
prediction. Moreover, from a general non-Gaussian analysis point of view, the introduction of an enfolded 
template might be interesting in itself, potentially providing complementary information in addition to the local 
and equilateral templates. The enfolded template proposed here was unfortunately only a marginal improvement over
the local and equilateral template, still being insensitive to the localized enhancement.
It would certainly be worthwhile to look for a more optimal enfolded template that could approach 
the theoretically maximum level of sensitivity to the localized enhancement, corresponding to a 
$\sqrt{|k_1 \eta_0|}=\sqrt{M/H}$ dependence of the corresponding fudge factor.

After adding a specific higher derivative term we surprisingly found that the localized nature of the enhancement is substituted by 
an overall enhancement of $(k_1 \eta_0)^2$ that can be absorbed directly into the non-Gaussian amplitude $f_{\mathrm{NL}}$. 
Even though (sub-leading) terms exist that are displaying a localized form of enhancement in the collinear limit, it turns out that 
the leading contribution is enhanced over the full comoving momentum triangle domain. In other words, the non-Gaussian signal is 
not of the enfolded type in this particular case and no enhancement sensitivity can be lost by the integration over all triangles. 
This dominant term to the three-point function is rapidly oscillating, causing the sign of the bispectrum to oscillate as well.
This oscillating sign feature implies that the leading contribution to the scalar product with the currently available templates 
is severely suppressed and are not sensitive to the full $(k_1 \eta_0)^2$ enhancement. Instead, subleading order $|k_1 \eta_0|$ 
terms also contribute to the scalar product with the local or equilateral template. The endresult is that the local and 
equilateral templates only probe a linear $k_1 \eta_0$ enhancement. The details of the 2d and 3d fudge factor are complicated
for the values of $|k_1 \eta_0|=M/H$ of interest, but the numerically determined 2d fudge factor for $M/H=10^3$ was used to put 
a constraint on the Bogoliubov parameter of order $10^{-2}$, close to the bound derived from the two-point power spectrum.
The main message however should be that improved templates, sensitive to the oscillatory nature of the dominant contribution to the bispectrum,
would considerably tighten these constraints. The oscillatory nature of the signal in momentum space suggests that specific, perhaps
observable, features could appear in 3d and 2d position space. In any case it would be worthwhile to generalize the range of available
non-Gaussian shapes that can be compared to the data, including oscillatory signals, which we hope to report on in the future. 
Theoretically at least, for an optimal template, this would lead to a limit on vacuum modifications 
orders of magnitude stronger than the bound obtained from the two-point power spectrum, which would be quite remarkable.

One important general conclusion supported by our results is that higher derivative corrections, which on general grounds
are always expected to be present, are extremely sensitive to departures from the standard Bunch-Davies 
vacuum state. Throughout this paper we assumed that the combination $|k_1 \eta_0|$ is independent of the actual comoving
momenta involved and equal to $M/H$, in the spirit of the New Physics Hypersurface approach to vacuum state modifications. 
The reason for this was scale invariance of the bispectrum, which we relied on to allow for comparison with the available (scale-invariant)
template shapes. Fixing $\eta_0$ instead, as one would do in a Boundary Effective Field Theory approach to vacuum state modifications, 
immediately results in a scale-dependent bispectrum. It would be interesting to study such scale-dependent scenarios, requiring
more general analysis tools \cite{Fergusson2008}, and determine to what extent (future) analysis of 3d large scale structure or 2d CMB data can constrain 
bispectrum departures from scale-invariance. As reported, the bispectrum or three-point function is extremely sensitive to initial-state modifications 
in the presence of a higher derivative operator, and there is no reason to think this could not similarly be true for all higher $n$-point functions. 
A more general perturbative analysis, including higher $n$-point functions, might lead to a hierarchy of 
(theoretical) constraints on vacuum state modifications, perhaps pointing to the standard Bunch-Davies 
state as the only consistent possibility in practice. We hope that ongoing future work in this direction can 
further help us understand and identify the phenomenological and theoretical constraints on the vacuum 
state ambiguity.

\vspace{0.5cm}

{\large{\bf{Acknowledgments}}}

We thank Benjamin Wandelt, Andrew Tolley and Paolo Creminelli for very useful discussions. PDM was
supported by the Netherlands Organization for Scientific Research (NWO), NWO-toptalent grant 021.001.040. 
The research of JPvdS is financially supported by Foundation of Fundamental Research on Matter (FOM)
grant 06PR2510. PDM and JPvdS were supported in part by a van Gogh Collaboration Grant VGP 63-254 from 
the Netherlands Organisation for Scientific Research (NWO) and PSC was supported by the van Gogh
program grant N-18150RG of the ``Minist\`ere des Affaires Etrang\`eres et Europ\'eennes''.
\vspace{0.5cm}

\appendix

\section{Enfolded triangles}

We have shown in section 2 that non-Gaussianities due to initial-state modifications for a canonical single inflaton action
enhances in the collinear limit, corresponding to enfolded or squashed triangles. Though the precise shape does not overlap perfectly, here we 
will translate our proposed template to an estimator to measure the associated amplitude $f_{\mathrm{NL}}^{\mathrm{enf}}$.

The estimators for different shapes of non-Gaussianity can be written
as follows \cite{Komatsu2008}
\begin{eqnarray*}
f_{\mathrm{NL}}^{\mathrm{local}} & = & (F^{-1})_{11}S_{1}+(F^{-1})_{12}S_{2}+(F_{}^{-1})_{13}S_{3}+(F^{-1})_{14}S_{4}\\
f_{\mathrm{NL}}^{\mathrm{equil}} & = & (F^{-1})_{22}S_{2}+(F^{-1})_{21}S_{1}+(F_{}^{-1})_{23}S_{3}+(F^{-1})_{24}S_{4}\\
f_{\mathrm{NL}}^{\mathrm{enf}} & = & (F^{-1})_{33}S_{3}+(F^{-1})_{31}S_{1}+(F_{}^{-1})_{32}S_{2}+(F^{-1})_{34}S_{4}\\
b_{src} & = & (F^{-1})_{44}S_{4}+(F^{-1})_{41}S_{1}+(F_{}^{-1})_{42}S_{2}+(F^{-1})_{43}S_{3}.\end{eqnarray*}
Here $F_{ij}$ represents the Fisher matrix and is inversely proportional
to the covariance, the overlap, between two bispectra. In case
one has a Gaussian likelihood and only one parameter to fit, the Fisher
matrix is equal to the inverse variance; $F_{ij}=1/\sigma_{\alpha}^{2}$,
with $\alpha$ the fitting parameter. Here it is given by\[
F_{ij}\equiv\sum_{2\leq l_{1}\leq l_{2}\leq l_{3}}\frac{B_{l_{1}l_{2}l_{3}}^{(i)}B_{l_{1}l_{2}l_{3}}^{(j)}}{\tilde{C}_{l_{1}}\tilde{C}_{l_{2}}\tilde{C}_{l_{3}}},\]
which is practically equal to eq. \eqref{eq:BB}. 
Once more, the $B_{l_{1}l_{2}l_{3}}^{(i)}$ are the theoretical bispectra of
the various non-Gaussian shapes. $\tilde{C}_{l}$ represents the
total angular power spectrum, which contains both the CMB signal and
the noise, i.e. $\tilde{C}_{l}=C_{l}^{cmb}b_{l}^{2}+N_{l}$. The $b_{l}$
is the beam transfer function, which is instrument dependent. If the
beam is Gaussian it has the form $b_{l}\propto\mathrm{exp}[-l^{2}\sigma_{b}^{2}]$,
where $\sigma_{b}=0.425\mathrm{FWHM}$. 

The estimators for the point sources, the local and the equilateral are not significantly modified when
multiple parameters are fit simultaneously \cite{Komatsu2008}, and from the calculations of the various cosine in the paper we can assume
this should hold for our template proposal, although to a lesser extent due to the small overlap between the local and enfolded shape. In the assumptions we can neglect this overlap, the Fisher matrix only has diagonal terms. Therefore
\[
f_{\mathrm{NL}}^{\mathrm{local}}=S_{1}/F_{11},\;\;\; f_{\mathrm{NL}}^{\mathrm{equil}}=S_{2}/F_{22},\;\;\; f_{\mathrm{NL}}^{\mathrm{enf}}=S_{3}/F_{33},\;\;\; b_{src}=S_{4}/F_{44}.\]
Note that one can now directly compute the pre-factor $\propto F_{ij}^{-1}$,
without first inverting the full Fisher matrix. 

In this paper we have proposed a template for the enfolded or squashed triangles, that should measure $f^\mathrm{enf}_{\mathrm{NL}}$
\begin{eqnarray}
F(k_{1},k_{2},k_{3}) & = & 6\Delta_{\Phi}^{2}\left[\frac{1}{k_{1}^{3}k_{2}^{3}}+2\;\mathrm{perm}+\frac{3}{k_{1}^{2}k_{2}^{2}k_{3}^{2}}\right.\nonumber \\
&  & \left.-\left(\frac{1}{k_{1}k_{2}^{2}k_{3}^{3}}+5\;\mathrm{perm}\right)\right]. \ \label{eq:enftemplate}\end{eqnarray}
Note that if we took a slightly different template, e.g., choosing a 4 instead of a 3 in the equation above, the following will still hold, 
and one simply needs to replace this 3 with a 4 from now on. 

To quantify to what extent the enfolded template gives complementary 
information once applied to the data, as compared to local and equilateral templates, 
we should consider the scalar product between the different templates. 
As pointed out in the main text, the starting point for deriving the most optimal enfolded 
template shape function is $F^{\mathrm{enf}}=-F^{\mathrm{equil}}+c/k_1^2k_2^2k_3^2$, in terms
of a general parameter $c$. We will plot the cosine of this template distribution with the equilateral template 
as a function of $c$, with $c$ running from $0$ to $4$. The result is shown in figure \ref{fig:enfeq}.
If we demand the template to have a definite (positive) sign, one should really only consider $c \geq 1$. 
In that case it should be clear that the optimal value, i.e. the smallest cosine equal to 
$\mathrm{Cos}(F^\mathrm{enf},F^\mathrm{equil})=0.49$ , is achieved for $c=1$, as claimed in section 5. 
For the cosine with the local template we find that it is more or less independent of $c$, as exemplified
by the fact that $\mathrm{Cos}(F^\mathrm{enf},F^\mathrm{local})_{c=3}/\mathrm{Cos} (F^\mathrm{enf},F^\mathrm{local})_{c=1}\simeq 1.06$ 
and growing ever slower for larger $c$.
For the cosine with the local template we find, for $c=1$, that $\mathrm{Cos}(F^\mathrm{enf},F^\mathrm{local})=0.67$.
This is quite large, which we should have expected since we can imagine the local template to be a special case of the 
factorized enfolded distribution, for which only the endpoints of the line $x_2+x_3=1$ are maximal, versus the whole line 
for the enfolded template. For completeness let us also mention the cosine between the local and equilateral template
$\mathrm{Cos}(F^\mathrm{equil},F^\mathrm{local})=0.41$.

\FIGURE{
\includegraphics[width=130mm]{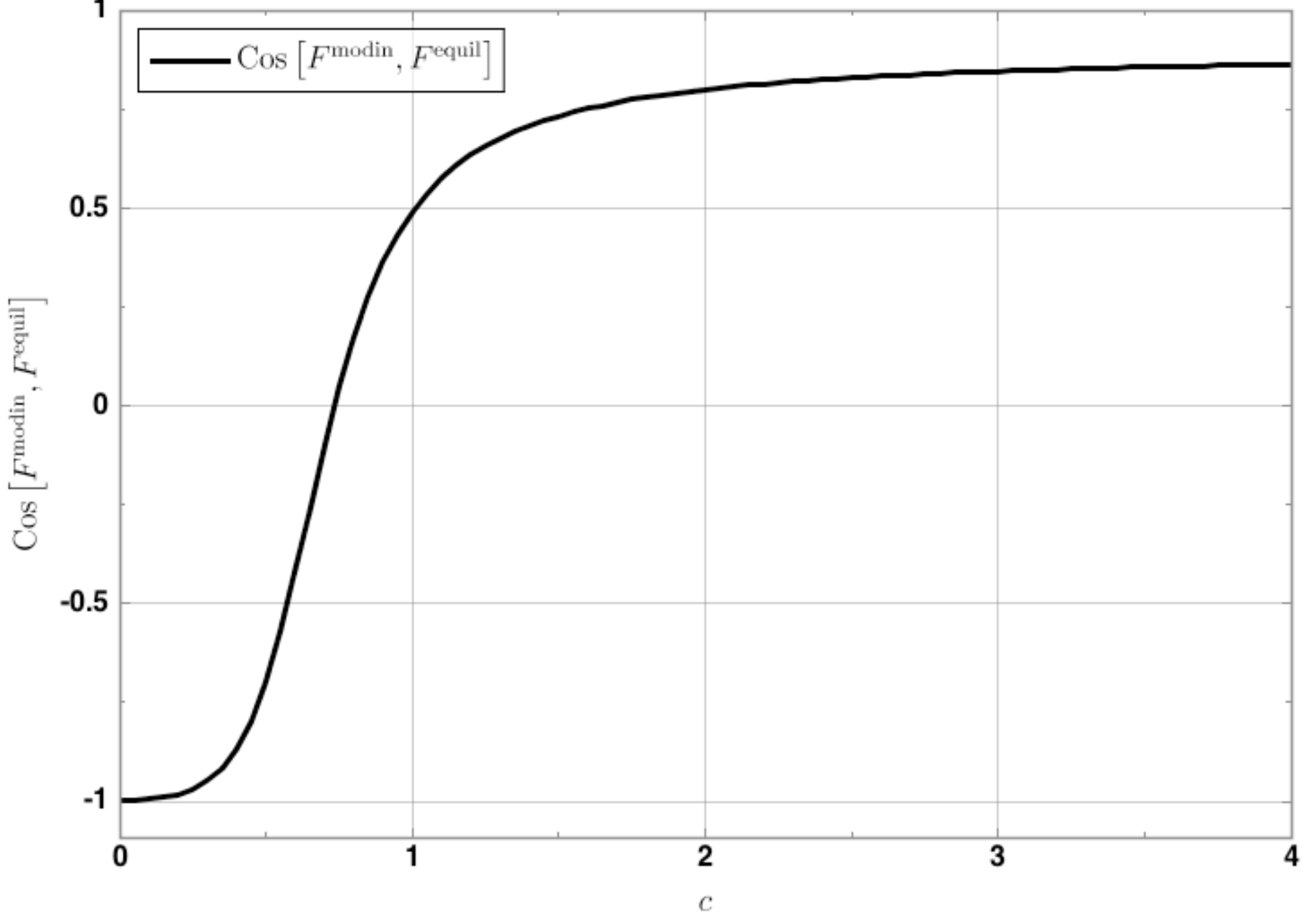} 
\caption{The plot shows how the $\mathrm{Cos}(F^\mathrm{enf},F^\mathrm{equil})$ changes as a function of $c$. The smaller the value of the cosine, 
the more distinct the two shapes are. As expected for $c=0$ the cosine is -1, i.e.
$F^{\mathrm{enf}}=-F^{\mathrm{equil}}$. Note that for 
$0<c<1$ the value of the cosine crosses zero. However these values of $c$ can not be used, because the sign of the three-point function should
be definite. For $0 < c<1$ this is not the case and the reason for getting a smaller value for the cosine is due to cancellations between positive and negative parts of the shape function.}
\label{fig:enfeq}
}

Any deviations from scale invariance can simply be taken into account
by replacing power of $n$ with $n-(n/3)(n_{s}-1)$, with $n_{s}$
the spectral index. If one forgets about the pre-factors
and divides out the $k_{1}$ dependence, the shape can plotted as
a function of $k_{2}/k_{1}$ and $k_{3}/k_{1}$. This is shown in
figure \ref{fig:enfoldedtemplate}.

Indeed the template shape maximizes when $k_{1}=k_{2}+k_{3}$. However
it does not blow up at this limit, which is the case when the denominator
is proportional to $k_{1}-k_{2}-k_{3}$. Such behavior can (possibly)
not be mimicked, using factorizable templates, Therefore such a denominator would not be allowed, since a function
with such a denominator can not be split up into functions of individual
comoving momenta $k_{1},$ $k_{2}$ and $k_{3}$. 

There exists another approach \cite{Smith2006}, in which such
a denominator is written as follows\[
\frac{1}{(k_{1}-k_{2}-k_{3})^{2}}=\int_{0}^{\infty}te^{-t(k_{1}-k_{2}-k_{3})}dt.\]
Now one has an integral over exponentials, which are factorizable. For eq. \eqref{fmodin}
this would imply 
\begin{eqnarray}
F^{\mathrm{modin}}(k_{1},k_{2},k_{3}) & \propto & \frac{1}{k_1 k_2 k_3}\sum_j\int_0^{\eta_0}\frac{\mathrm{sin}(\tilde{k}_j t)}{k_j^2} dt.
\end{eqnarray}
Obviously this introduces another integration, increasing computational
time one is trying to win by factorizing. In \cite{Smith2006} it was shown that the double integration
can be done rather quickly for an equilateral template. In our case, this might not be possible because 
of the large number of oscillations in the cosine, which we expect to require a large number of
quadrature points, when one replaces the integral over $t$ by a weighted sum. For
now, we will focus on the enfolded template shape of eq. \eqref{eq:enftemplate} and leave the 
investigation of the method above for future work.

One can define the following maps\begin{eqnarray}
\alpha_{l}(r) & = & \frac{2}{\pi}\int k^{2}dk\Delta_{l}(k)j_{l}(kr)\label{eq:alpha}\\
\beta_{l}(r) & = & \frac{2}{\pi}\int k^{2}dkP_{\Phi}\Delta_{l}(k)j_{l}(kr)\label{eq:beta}\\
\gamma_{l}(r) & = & \frac{2}{\pi}\int k^{2}dkP_{\Phi}^{1/3}\Delta_{l}(k)j_{l}(kr)\label{eq:gamma}\\
\delta(r) & = & \frac{2}{\pi}\int k^{2}dkP_{\Phi}^{2/3}\Delta_{l}(k)j_{l}(kr)\label{eq:delta}\end{eqnarray}
Here $\Delta_{l}(k)$ is the photon transfer function, introduced in section  6, which is used to compute
the (theoretical) angular power spectrum $C_{l}=(2/\pi)\int k^{2}dkP_{\Phi}(k)\Delta_{l}^{2}$.
$P_{\phi}(k)$ is the primordial power spectrum $P_{\Phi}(k)\propto k^{n_{s}-1}/k^{3}$.
It can be seen that all maps have a different primordial power spectrum
dependence (and consequently different dimension). These maps are
required to set up an estimator that has the same $k$ dependence
as the template \eqref{eq:enftemplate}. 

Using eq. \eqref{eq:alpha} through \eqref{eq:delta} one can
construct 4 `filtered' maps (recall that $\Delta T(\hat{n})=\sum_{lm}a_{lm}Y_{lm}(\hat{n})$)\begin{eqnarray}
A(\hat{n},r) & = & \sum_{l=2}^{l_{max}}\sum_{m=-l}^{l}\alpha_{l}(r)\frac{b_{l}}{\tilde{C}_{l}}a_{lm}Y_{lm}(\hat{n}),\label{eq:filter A}\\
B(\hat{n},r) & = & \sum_{l=2}^{l_{max}}\sum_{m=-l}^{l}\beta_{l}(r)\frac{b_{l}}{\tilde{C}_{l}}a_{lm}Y_{lm}(\hat{n}),\label{eq:filter B}\\
C(\hat{n},r) & = & \sum_{l=2}^{l_{max}}\sum_{m=-l}^{l}\gamma_{l}(r)\frac{b_{l}}{\tilde{C}_{l}}a_{lm}Y_{lm}(\hat{n}),\label{eq:filter C}\\
D(\hat{n},r) & = & \sum_{l=2}^{l_{max}}\sum_{m=-l}^{l}\delta_{l}(r)\frac{b_{l}}{\tilde{C}_{l}}a_{lm}Y_{lm}(\hat{n}).\label{eq:filter D}\end{eqnarray}
The sum over $l$ runs from 2 to $l_{max}$, since the monopole and
the dipole are hard/impossible to measure and WMAP (or any other instrument
for that matter) can only measure up to a certain $l_{max}$ based
on the instrument's technical limitations. Now we need to set up a
bispectrum that has the `same' comoving momentum dependence as the
template. For reasons that will become clear later, it is convenient
to define the bispectrum related to local shape ($f_{\mathrm{NL}}^{\mathrm{local}}$).
The local shape is local in real space and its template is exact,
that is, the theoretical shape is equivalent to a factorized template
\begin{eqnarray}
F(k_{1},k_{2},k_{3})&=&f_{\mathrm{NL}}^{\mathrm{local}}\Delta_{\Phi}^{2}2\left(\frac{1}{k_{1}^{3}k_{2}^{3}}+\frac{1}{k_{2}^{3}k_{3}^{3}}+\frac{1}{k_{3}^{3}k_{1}^{3}}\right).
\end{eqnarray}
It can be seen from the template that the shape is proportional to
a product of power spectra ($n_{s}=1$). We can thus use cyclic product
of the angular maps \eqref{eq:alpha} and \eqref{eq:beta}. The (angular
averaged) bispectrum can be written as 
\begin{eqnarray}
B_{l_{1}l_{2}l_{3}}^{\mathrm{local}}&=&2I_{l_{1}l_{2}l_{3}}\int_{0}^{\infty}r^{2}dr\left[\alpha_{l_{1}}(r)\beta_{l_{2}}(r)\beta_{l_{3}}(r)+\mathrm{cycl.\; perm}\right].
\end{eqnarray}
One integrates over the comoving distance. The sampling rate (in $r$
space) depends on the behavior of the transfer function $\Delta_{l}$.
In addition, $I_{l_{1}l_{2}l_{3}}$ is known as the Gaunt factor and
is a product of the solid angle integration of (angular averaged bispectrum
remember) the $Y_{lm}$. It is given by (eq. \eqref{eq:Blll})
\begin{eqnarray}
I_{l_{1}l_{2}l_{3}}&=&\sqrt{\frac{(2l_{1}+1)(2l_{2}+1)(2l_{3}+1)}{4\pi}}\left(\begin{array}{ccc}
l_{1} & l_{2} & l_{3}\\
0 & 0 & 0\end{array}\right).
\end{eqnarray}
The first term in the enfolded shape is equivalent to the local shape,
so we can use $B_{l_{1}l_{2}l_{3}}^{\mathrm{local}}$ to express (partly)
$B_{l_{1}l_{2}l_{3}}^{\mathrm{enf}}$. The rest is obtained via carefully
combining products of the angular maps \eqref{eq:alpha} through \eqref{eq:delta}\[
B_{l_{1}l_{2}l_{3}}^{\mathrm{enf}}=3B_{l_{1}l_{2}l_{3}}^{\mathrm{local}}+6I_{l_{1}l_{2}l_{3}}\int r^{2}dr\left[-\left(\beta_{l_{1}}(r)\gamma_{l_{2}}(r)\delta_{l_{3}}(r)+\mathrm{cycl.\; perm}\right)+3\delta_{l_{1}}(r)\delta_{l_{2}}(r)\delta_{l_{3}}(r)\right]\]
Now one can easily set up the skewness estimator as explained in \cite{Yadav2007a, Yadav2007b, Komatsu2008} \[
S^{\mathrm{enf}}=S_{\mathrm{prim}}+S_{\mathrm{prim}}^{\mathrm{linear}},\]
where the first term is simply the term that represents the shape of
the bispectrum (the cubic term, cubic in the filtered maps), while
the second, the linear term (linear in the filtered maps), is added
to minimize the effect caused by the inhomogeneous noise that breaks
rotational invariance ($\hat{C}_{l}$ will have off-diagonal terms).
The linear term should be constructed such that it minimizes the variance
of the estimator. If the linear term
is constructed as follows, this can indeed by achieved. One first has to derive the filtered maps
$A$, $B$, $C$ and $D$ of eq. \eqref{eq:filter A}, \eqref{eq:filter B},
\eqref{eq:filter C} and \eqref{eq:filter D} that can be used to
set up the cubic statistic estimator, $S_{\mathrm{prim}}$. Subsequently
one takes the Monte Carlo average, $\langle S_{\mathrm{prim}}\rangle_{\mathrm{MC}}$.
Now let us suppose that $S_{\mathrm{prim}}$ is constructed out of
3 filtered maps $A$, $B$ and $C$. One can apply Wick's theorem
to rewrite the average of a cubic product $\langle ABC\rangle_{\mathrm{MC}}=\langle A\rangle_{\mathrm{MC}}\langle BC\rangle_{\mathrm{MC}}+\langle B\rangle_{\mathrm{MC}}\langle AC\rangle_{\mathrm{MC}}+\langle C\rangle_{\mathrm{MC}}\langle AB\rangle_{\mathrm{MC}}$.
Next, remove the $\mathrm{MC}$ average from the single maps and replace
the maps within brackets with simulated maps. Our linear estimator
becomes: $A\langle B_{\mathrm{sim}}C_{\mathrm{sim}}\rangle_{\mathrm{MC}}+B\langle A_{\mathrm{sim}}C_{\mathrm{sim}}\rangle_{\mathrm{MC}}+C\langle A_{\mathrm{sim}}B_{\mathrm{sim}}\rangle_{\mathrm{MC}}$.
If we apply this trick and apply weighting functions we get \begin{eqnarray}
S^{\mathrm{local}} & \equiv & 4\pi\int r^{2}dr\int\frac{d^{2}\hat{n}}{w_{3}}\left(A(\hat{n},r)B^{2}(\hat{n},r)\right. \nonumber \\
&  & \left.-2B(\hat{n},r)\langle A_{\mathrm{sim}}(\hat{n},r)B_{\mathrm{sim}}(\hat{n},r)\rangle_{\mathrm{MC}}-A(\hat{n},r)\langle B_{\mathrm{sim}}^{2}(\hat{n},r)\rangle_{\mathrm{MC}}\right),\end{eqnarray}
for the local estimator and \begin{eqnarray}
S^{\mathrm{enf}} & \equiv & 3S^{\mathrm{local}}\nonumber\\ 
&&+24\pi\int r^{2}dr\int\frac{d^{2}\hat{n}}{w_{3}}\left[\left(-B(\hat{n},r)C(\hat{n},r)D(\hat{n},r)+B(\hat{n},r)\langle C_{\mathrm{sim}}(\hat{n},r)D_{\mathrm{sim}}(\hat{n},r)\rangle_{\mathrm{MC}}\right.\right.\nonumber \\
&  & \left.+C(\hat{n},r)\langle B_{\mathrm{sim}}(\hat{n},r)D_{\mathrm{sim}}(\hat{n},r)\rangle_{\mathrm{MC}}+D(\hat{n},r)\langle B_{\mathrm{sim}}(\hat{n},r)C_{\mathrm{sim}}(\hat{n},r)\rangle_{\mathrm{MC}}\right) \nonumber \\
&  & \left.+\left(D^{3}(\hat{n},r)-4D(\hat{n},r)\langle D_{\mathrm{sim}}^{2}(\hat{n},r)\rangle_{\mathrm{MC}}\right)\right]\end{eqnarray}
for the enfolded estimator. Here $w_{3}$ is sum of the weighting
functions cubed\[
w_{3}=\int d^{2}\hat{n}W^{3}(\hat{n}).\]
The cube is a result of the fact that one has 3 $a_{lm}$'s in each
cubic product of the filtered maps in the skewness estimators. In
real space a mask is simply a multiplication (i.e. one can multiply
each $\Delta T(\hat{n})$ with either zero or one). However, this
becomes a convolution in Fourier space. Consequently we have an integral
over the solid angle $d^{2}\hat{n}$. If there is uniform weighting
(that is, each pixel is masked or not) $W(\hat{n})=M(\hat{n})$, the
mask function and the integral becomes \[
w_{3}=4\pi f_{sky}.\]
with $f_{sky}$ the sky cut (in fact it is the opposite, it represents
the percentage of sky that remains after masking). In the latest WMAP
analysis they have not used a uniform weighting, but a {}``combination
signal-plus-noise weight'', which turns out to be optimal for the
analysis of equilateral shaped bispectra, while the local shape is
barely affected by simply using the uniform weight. It should be checked
to what extent uniform weighting changes the estimates of the enfolded
shape, compared to the more advanced combined weighting used by WMAP team \cite{Komatsu2008}.

\section{Modified initial-state bispectra}

We start with eq. (3.11) in \cite{Holman2007}, from which we can extract
the interaction Hamiltonian from the canonical effective action minimally
coupled to gravity
\begin{equation}
H_{I}=-\frac{H}{M_{p}^{2}}\int d^{3}xa(\eta)^{3}\left(\frac{\dot{\phi}}{H}\right)^{4}\zeta'^{2}\partial^{-2}\zeta'.\label{Hinteract}
\end{equation}
As carefully explained in \cite{Holman2007} the three-point correlation
function $\langle\zeta_{k_{1}}\zeta_{k_{2}}\zeta_{k_{3}}\rangle$
can be written as (to first order in the interaction Hamiltonian $H_{I}$)
an integral over the free field correlator $\langle\zeta_{k_{1}}\zeta_{k_{2}}\zeta_{k_{3}}H_{I}(\eta)\rangle$,
where the free field correlator can be expanded in a product of two
point correlators (i.e., Wightman functions) via Wick's theorem. Consequently
it is straightforward to show that the three-point correlation function
in the case we consider the interaction Hamiltonian of eq. \eqref{Hinteract}
is given by\begin{eqnarray}
\langle\zeta_{k_{1}}\zeta_{k_{2}}\zeta_{k_{3}}\rangle & = & -i(2\pi)^{3}\delta^{(3)}\left(\sum\vec{k}_{i}\right)\frac{H}{M_{p}^{2}}\left(\frac{\dot{\phi}}{H}\right)^{4}\times\\
& &\int_{\eta_{)}}^{0}d\eta a^{3}(\eta)\frac{1}{k_{3}^{2}}\partial_{\eta}G_{k_{1}}^{>}(0,\eta)\partial_{\eta}G_{k_{2}}^{>}(0,\eta)\partial_{\eta}G_{k_{3}}^{>}(0,\eta) +\mathrm{perm+c.c,}\label{G1}\end{eqnarray}
as can be found in \cite{Holman2007}. Here the Wightman functions $G_{k}^{>}$
are defined as followed
\begin{equation}
\langle\zeta_{k_{1}}\zeta_{k_{2}}\rangle=(2\pi)^{3}\delta^{(3)}(\vec{k}_{1}+\vec{k}_{2})G_{k_{1}}^{>}(\eta,\eta').\label{G2}
\end{equation}
The Wightman functions can be found by solving the e.o.m. of the
inflaton field minimally coupled to gravity
\begin{equation}
G_{k}^{>}(\eta,\eta')=\frac{H^{2}}{\dot{\phi}^{2}}\frac{H^{2}}{2k^{3}}(1+ik\eta)(1-ik\eta')e^{-ik(\eta-\eta')}.
\end{equation}
Consequently we compute 
\begin{equation}
G_{k}^{>}(0,\eta)=\frac{H^{2}}{\dot{\phi}^{2}}\frac{H^{2}}{2k^{3}}(1-ik\eta)e^{ik\eta},
\end{equation}
and
\begin{equation}
\partial_{\eta}G_{k}^{>}(0,\eta)=-\frac{H^{2}}{\dot{\phi}^{2}}\frac{H}{2k}\frac{1}{a(\eta)}e^{ik\eta},
\end{equation} 
with $a(\eta)=1/\eta H$ during inflation in the assumption $\dot{H}\simeq0$. 

Next we can express $\langle\zeta_{k_{1}}\zeta_{k_{2}}\zeta_{k_{3}}\rangle$
in terms of these Wightman functions. However, we are interested in
the case where we are not in the BD vacuum state. To first order in
the Bogoliubov parameter $\beta_{k}$, what happens is that one of
the solutions to the equation of motion `picks up' a minus sign
in $k$. It is easy to incorporate this minus sign to find
\begin{equation}
\langle\zeta_{k_{1}}\zeta_{k_{2}}\zeta_{k_{3}}\rangle^{\mathrm{nBD}}=-i(2\pi)^{3}\delta^{(3)}\left(\sum\vec{k}_{i}\right)\frac{1}{M_{p}^{2}}\frac{2}{\prod(2k_{i}^{3})}\frac{H^{6}}{\dot{\phi}^{2}}\int_{\eta_{0}}^{0}d\eta\sum_{j}\beta_{k_{j}}^{*}\frac{3 k_{1}^{2}k_{2}^{2}k_{3}^{2}}{k_{j}^{2}}e^{i\tilde{k}_{j}\eta}+\mathrm{c.c.}\label{modin1}
\end{equation}
Note the factor of 6 is the result of the 6 possible permutations, while the sum is the result from implementing an
initial-state modification for all of these 6 permutations. $\tilde{k}_{j}=k_{t}-2k_{j}$,
with $k_{t}=k_{1}+k_{2}+k_{3}$. The integral can be easily performed
as well as adding the complex conjugate part. In steps\[
-i\times\int_{\eta_{0}}^{0}d\eta e^{i\tilde{k}_{j}\eta}+\mathrm{c.c.}=\frac{2(\mathrm{cos}(\tilde{k}_{j}\eta_{0})-1)}{\tilde{k}_{j}}\]
resulting in
\begin{equation}
\langle\zeta_{k_{1}}\zeta_{k_{2}}\zeta_{k_{3}}\rangle^{\mathrm{nBD}}=(2\pi)^{3}\delta^{(3)}\left(\sum\vec{k}_{i}\right)\frac{1}{M_{p}^{2}}\frac{4}{\prod(2k_{i}^{3})}\frac{H^{6}}{\dot{\phi}^{2}}\sum_{j}\frac{3 k_{1}^{2}k_{2}^{2}k_{3}^{2}}{k_{j}^{2}\tilde{k}_{j}}\mathcal{R}e(\beta_{k_{j}})\left(\mathrm{cos}(\tilde{k}_{j}\eta_{0})-1\right).
\end{equation}
This is the result we have used in eq. \eqref{modin2} of section 2.
In case we assume that the enhancement occurs when $\tilde{k}_{j}\rightarrow0$,
we can apply this limit to the above expression to find \begin{eqnarray*}
\langle\zeta_{k_{1}}\zeta_{k_{2}}\zeta_{k_{3}}\rangle^{\mathrm{nBD}} & = & -(2\pi)^{3}\delta^{(3)}\left(\sum\vec{k}_{i}\right)\frac{1}{M_{p}^{2}}\frac{4}{\prod(2k_{i}^{3})}\frac{H^{6}}{\dot{\phi}^{2}}\sum_{j}\frac{3 k_{1}^{2}k_{2}^{2}k_{3}^{2}}{k_{j}^{2}}\mathcal{R}e(\beta_{k_{j}})\\
&  & \times\frac{1}{2}\tilde{k}_{j}\eta_{0}^{2}+\mathcal{O}(\tilde{k}_{j}^{3}),\end{eqnarray*}
which is slightly different from the result found in \cite{Holman2007} since
they did not consider the limit of $x\rightarrow0$ in $\mathrm{cos}(x)/x$
correctly. Note that when $\tilde{k}_{j}=0$ this whole expression
actually vanishes, but does have a maximum nearby (i.e., $k_{max}\sim\eta_0^{-1}$).

Next we consider higher-order terms in the action of the form 
\begin{equation}
\mathcal{L}_{I}=\sqrt{-g}\frac{\lambda}{8M^{4}}((\nabla\Phi)(\nabla\Phi))^{2}.
\end{equation}
It is not hard to compute the corresponding interaction Hamiltonian
up to third order in the curvature field $\zeta$ (and $\zeta\simeq-(H/\dot{\phi})\delta\phi$,
where $\Phi=\phi(\eta)+\delta\phi(\eta,x)$). It can be shown that
\begin{equation}
H_{I}=-\frac{\lambda H}{2M^{4}}\int d^{3}xa(\eta)\left(\frac{\dot{\phi}}{H}\right)^{3}\zeta'\left(\zeta'^{2}-(\partial_{i}\zeta)^{2}\right).\label{HIHD}
\end{equation}
In similar fashion we can write down the three-point correlator in
terms of integrals over products of Wightman functions, while Fourier
transforming to $k$ space\begin{eqnarray}
\langle\zeta_{k_{1}}\zeta_{k_{2}}\zeta_{k_{3}}\rangle^{\mathrm{HD}} & = & -i(2\pi)^{3}\delta^{(3)}\left(\sum\vec{k}_{i}\right)\frac{\lambda}{2HM^{4}}\left(\frac{\dot{\phi}}{H}\right)^{3}\int d\eta a(\eta)\nonumber\\
\left(\partial_{\eta}G_{k_{1}}^{>}(0,\eta)\partial_{\eta}G_{k_{2}}^{>}(0,\eta)\partial_{\eta}G_{k_{3}}^{>}(0,\eta)\right.
&+&\left. \vec{k}_{1}\cdot\vec{k}_{2}G_{k_{1}}^{>}(0,\eta)G_{k_{2}}^{>}(0,\eta)\partial_{\eta}G_{k_{3}}^{>}(0,\eta)+\mathrm{perm}\right),\nonumber \end{eqnarray}
and its complex conjugate.
Here the dot product comes from the Fourier transform of two partial
spatial derivatives in the interaction Hamiltonian of eq. \eqref{HIHD}. This
is a much `larger' three-point correlator, so let us compute it in
such a way that we do not lose track of all different components.
The best approach is to first compute all the different terms in the
integral and then group these in proportionality to $\eta$ (i.e.
$\propto\eta^{0},\eta$ and $\eta^{2}$) . In addition we can compute
the whole `pre-factor' independently. It is not hard to show that it
is given by 
\begin{equation}
P=(2\pi)^{3}\delta^{(3)}\left(\sum\vec{k}_{i}\right)\frac{\lambda}{M^{4}}\frac{1}{\prod(2k_{i}^{3})}\frac{H^{8}}{\dot{\phi}^{2}},
\end{equation}
and therefore 
\begin{eqnarray}
\langle\zeta_{k_{1}}\zeta_{k_{2}}\zeta_{k_{3}}\rangle_{\mathrm{nBD}}^{\mathrm{HD}} & = & -iP\int_{\eta_{0}}^{0}d\eta\sum_{j}\beta_{k_{j}}^{*}\left[3\eta^{2}k_{1}^{2}k_{2}^{2}k_{3}^{2}\right.\nonumber \\
&  & \left.+\vec{k}_{j}\cdot\vec{k}_{j+1}k_{j+2}^{2}(1+ik_{j}\eta)(1-ik_{j+1}\eta)\right.\nonumber \\ 
&  & \left.+\vec{k}_{j}\cdot\vec{k}_{j+2}k_{j+1}^{2}(1+ik_{j}\eta)(1-ik_{j+2}\eta)\right.\nonumber \\ 
&  & \left.+\vec{k}_{j+1}\cdot\vec{k}_{j+2}k_{j}^{2}(1-ik_{j+1}\eta)(1-ik_{j+2}\eta)\right]e^{i\tilde{k}_{j}\eta},
\end{eqnarray}
where the $j$'s are cyclic in $1$, $2$ and $3$. 

Let us first rewrite the dot product using the triangle vector constraint
$\vec{k}_{1}+\vec{k}_{2}+\vec{k}_{3}=0$. Using this constraint we
can deduce that \[
\vec{k}_{j}\cdot\vec{k}_{j+1}=\frac{1}{2}(k_{j+2}^{2}-k_{j}^{2}-k_{j+1}^{2}).\]
If we write the integral as $\int_{\eta_{0}}^{0}d\eta\sum_{j}\beta_{k_{j}}^{*}S(k_{1},k_{2},k_{3},\eta)e^{i\tilde{k}_{j}\eta}$
we obtain \begin{eqnarray}
S(k_{1},k_{2},k_{3},\eta) & = & +\frac{1}{2}\left[(k_{j+2}^{2}-k_{j}^{2}-k_{j+1}^{2})k_{j+2}^{2}+k_{j+1}^{2}-k_{j}^{2}-k_{j+2}^{2})k_{j+1}^{2}\right.\nonumber \\
&  & \left.+(k_{j}^{2}-k_{j+1}^{2}-k_{j+2}^{2})k_{j}^{2}\right]+\frac{i\eta}{2}\left[(k_{j+2}^{2}-k_{j}^{2}-k_{J+1}^{2})k_{j+2}^{2}(k_{j}-k_{j+1})\right. \nonumber \\
&  & \left.+(k_{j+1}^{2}-k_{j}^{2}-k_{j+2}^{2})k_{j+1}^{2}(k_{j}-k_{j+2})\right.\nonumber\\
& & \left.-(k_{j}^{2}-k_{j+1}^{2}-k_{J+2}^{2})k_{j}^{2}(k_{j+1}+k_{j+2})\right] \nonumber \\
&  & +\frac{\eta^{2}}{2}\left[k_{j}k_{j+1}k_{j+2}\left((k_{j+2}(k_{j+2}^{2}-k_{j}^{2}-k_{j+1}^{2})+k_{j+1}(k_{j+1}^{2}-k_{j}^{2}-k_{j+2}^{2})\right.\right.\nonumber \\
&  & \left.\left.-k_{j}(k_{j}^{2}-k_{j+1}^{2}-k_{j+2}^{2})\right)+6k_{1}^{2}k_{2}^{2}k_{3}^{2}\right].\end{eqnarray}
This can be rewritten as \begin{eqnarray}
S(k_{1},k_{2},k_{3},\eta) & = & -\frac{1}{2}k_{t}\prod_{i}\tilde{k}_{i}+\frac{i\eta}{2}\left[\tilde{k}_{j}\left(2k_{j+1}k_{j+2}(k_{j+1}+k_{j+2})^{2}\right.\right.\nonumber \\
&  & \left.\left.-k_{t}(\tilde{k}_{j}(k_{j+1}^{2}+k_{j+2}^{2})+(k_{j+1}+k_{j+2})(2k_{j+1}^{2}+3k_{j+1}k_{j+2}+2k_{j+2}^{2})\right.\right.\nonumber \\
&  & \left.\left.-\tilde{k}_{j}(k_{j+1}+k_{j+2})k_{t}\right)\right.\nonumber\\
&  & \left. -4k_{j+1}k_{j+2}(k_{j+1}+k_{j+2})(k_{j+1}^{2}+k_{j+2}^{2}+k_{j+1}k_{j+2})\right] \nonumber \\
&  & +\frac{\eta^{2}}{2}\left[\tilde{k}_{j}\prod_{i}k_{i}(k_{t}^{2}-4k_{j+1}k_{j+2}\right],\label{SHD}\end{eqnarray}
which is very similar to eq. (3.30)  in \cite{Holman2007} except for the term linear
in $\eta$, for which we find additional terms. Since the integral
runs over conformal time $\eta$, all that is left is to compute the different
integrals in terms of $\eta$. Again, it is useful to pre-compute the following integrals \begin{eqnarray*}
-i\times\int_{\eta_{0}}^{0}d\eta e^{i\tilde{k}_{j}\eta}+\mathrm{c.c.} & = & \frac{2(\mathrm{cos}(\tilde{k}_{j}\eta_{0})-1)}{\tilde{k}_{j}}\\
-i\times\int_{\eta_{0}}^{0}d\eta(i\eta)e^{i\tilde{k}_{j}\eta}+\mathrm{c.c.} & = & \frac{-2\eta_{0}\mathrm{sin}(\tilde{k}_{j}\eta_{0})}{\tilde{k}_{j}}+\frac{2(1-\mathrm{cos}(\tilde{k}_{j}\eta_{0}))}{\tilde{k}_{j}^{2}}\\
-i\times\int_{\eta_{0}}^{0}d\eta\eta^{2}e^{i\tilde{k}_{j}\eta}+\mathrm{c.c.} & = & \frac{2\eta_{0}^{2}\mathrm{cos}(\tilde{k}_{j}\eta_{0})}{\tilde{k}_{j}}-\frac{4\eta_{0}\mathrm{sin}(\tilde{k}_{j}\eta_{0})}{\tilde{k}_{j}^{2}}+\frac{4(1-\mathrm{cos}(\tilde{k}_{j}\eta_{0}))}{\tilde{k}_{j}^{3}}.\end{eqnarray*}
Since we now have terms that are proportional to $\tilde{k}_{j}^{3}$
one would expect some terms to diverge in case $\tilde{k}_{j}\rightarrow0$.
However, this proportionality only appears in the last term of eq. \eqref{SHD},
which is proportional to $\tilde{k}_{j}$ on itself. Consequently we
lose a factor of $\tilde{k}_{j}$ after multiplication, just enough
to make that term finite for $\tilde{k}_{j}\rightarrow0$. Similar
argumentation can be applied to the other terms, once we realize that
the limits of $\mathrm{sin}(x)/x$ and $(1-\mathrm{cos}(x))/x^{2}$
are finite in the limit $x\rightarrow0$. The three-point correlator therefore becomes\begin{eqnarray}
\langle\zeta_{k_{1}}\zeta_{k_{2}}\zeta_{k_{3}}\rangle_{\mathrm{nBD}}^{\mathrm{HD}} & = & (2\pi)^{3}\delta^{(3)}\left(\sum\vec{k}_{i}\right)\frac{\lambda}{M^{4}}\frac{1}{\prod(2k_{i}^{3})}\frac{H^{8}}{\dot{\phi}^{2}}\sum_{j}2\mathcal{R}e(\beta_{k_{j}})\nonumber \\
&  & \times\left[\frac{(1-\mathrm{cos}(\tilde{k}_{j}\eta_{0}))}{\tilde{k}_{j}}\left(-k_{t}\prod_{i}\tilde{k}_{i}+\mathcal{A}_{1}(k)+\frac{1}{\tilde{k}_{j}}\left(\mathcal{A}_{2}(k)+2\mathcal{B}(k)\right)\right)\right.\nonumber \\
&  & \left.-\frac{\eta_{0}\mathrm{sin}(\tilde{k}_{j}\eta_{0})}{\tilde{k}_{j}}\left(\tilde{k}_{j}\mathcal{A}_{1}+\mathcal{A}_{2}(k)+2\mathcal{B}(k)\right)+\eta_{0}^{2}\mathrm{cos}(\tilde{k}_{j}\eta_{0})\mathcal{B}(k)\right],\end{eqnarray}
where \begin{eqnarray*}
\mathcal{A}_{1}(k_{j},k_{j+1},k_{j+2}) & = & 2k_{j+1}k_{j+2}(k_{j+1}+k_{j+2})^{2}-k_{t}\left[\tilde{k}_{j}(k_{j+1}^{2}+k_{j+2}^{2})\right.\\
&  & \left.+(k_{j+1}+k_{j+2})(2k_{j+1}^{2}+3k_{j+1}k_{j+2}+2k_{j+2}^{2})\right]-\tilde{k}_{j}(k_{j+1}+k_{j+2})k_{t}^{2}\\
\mathcal{A}_{2}(k_{j},k_{j+1},k_{j+2}) & = & -4k_{j+1}k_{j+2}(k_{j+1}+k_{j+2})(k_{j+1}^{2}+k_{j+2}^{2}+k_{j+1}k_{j+2})\\
\mathcal{B}(k_{j},k_{j+1},k_{j+2}) & = & \prod_{i}k_{i}(k_{t}^{2}-4k_{j+1}k_{j+2}).\end{eqnarray*}
Unlike in the simple case, and aside the comoving dependence of the
nominator, this three-point correlator has terms that are proportional
to $(1-\mathrm{cos}(\tilde{k}_{j}\eta_{0}))/\tilde{k}_{j}^{2}$ and
$\mathrm{sin}(\tilde{k}_{j}\eta_{0})/\tilde{k}_{j}$. Therefore we
can take the exact limit $\tilde{k}_{j}=0$, and have a non-vanishing
result. The easiest way to see what happens is to look at the integrals.
In the limit $\tilde{k}_{j}=0$ the first integral and the last integral
vanish. Consequently, we can write \begin{eqnarray*}
\langle\zeta_{k_{1}}\zeta_{k_{2}}\zeta_{k_{3}}\rangle_{\mathrm{\tilde{k}_{j}=0}}^{\mathrm{HD}} & = & -(2\pi)^{3}\delta^{(3)}\left(\sum\vec{k}_{i}\right)\frac{\lambda}{M^{4}}\frac{1}{\prod(2k_{i}^{3})}\frac{H^{8}}{\dot{\phi}^{2}}\sum_{j}2\mathcal{R}e(\beta_{k_{j}})\\
&  & \times\frac{\eta_{0}^{2}}{2}\mathcal{A}_{2}(k_{j},k_{j+1},k_{j+2})\\
& = & -(2\pi)^{3}\delta^{(3)}\left(\sum\vec{k}_{i}\right)\frac{\lambda}{M^{4}}\frac{1}{\prod(2k_{i}^{3})}\frac{H^{8}}{\dot{\phi}^{2}}\sum_{j}2\mathcal{R}e(\beta_{k_{j}})\\
&  & \times\frac{\eta_{0}^{2}}{2}\left(-4k_{j+1}k_{j+2}(k_{j+1}+k_{j+2})(k_{j+1}^{2}+k_{j+2}^{2}+k_{j+1}k_{j+2})\right),\end{eqnarray*}
which is equivalent, up to a minus sign, to what the authors of \cite{Holman2007} found. However, we have shown here and have argued in section $5$ that 
this limit does not represent a maximum.
$ $


\begin{thebibliography}{10}

\bibitem{Kofman1991}
L.~A.~Kofman,
``Primordial perturbations from inflation,''
Phys.\ Scripta {\bf T36} (1991) 108.

\bibitem{Komatsu2002}
E.~Komatsu,
``The Pursuit of Non-Gaussian Fluctuations in the Cosmic Microwave
Background,''
arXiv:astro-ph/0206039.


\bibitem{ghostinfl_ng}
N.~Arkani-Hamed, P.~Creminelli, S.~Mukohyama and M.~Zaldarriaga,
``Ghost inflation,''
JCAP {\bf 0404}, 001 (2004)
[arXiv:hep-th/0312100].

L.~Senatore,
``Tilted ghost inflation,''
Phys.\ Rev.\  D {\bf 71}, 043512 (2005)
[arXiv:astro-ph/0406187].


\bibitem{exotic_ng}

I.~G.~Moss and C.~Xiong,
``Non-Gaussianity in fluctuations from warm inflation,''
JCAP {\bf 0704}, 007 (2007)
[arXiv:astro-ph/0701302].

F.~Bernardeau and T.~Brunier,
``Non-Gaussianities in extended D-term inflation,''
Phys.\ Rev.\  D {\bf 76}, 043526 (2007)
[arXiv:0705.2501 [hep-ph]].

N.~Barnaby and J.~M.~Cline,
``Large NonGaussianity from Nonlocal Inflation,''
JCAP {\bf 0707}, 017 (2007)
[arXiv:0704.3426 [hep-th]].

N.~Barnaby and J.~M.~Cline,
``Predictions for NonGaussianity from Nonlocal Inflation,''
arXiv:0802.3218 [hep-th].


\bibitem{fnl_local}

J. R. ~Bond and D. S. ~Salopek, 
Phys. \ Rev.\ D {\bf42} 3936 (1990).

T.~Falk, R.~Rangarajan and M.~Srednicki,
``The Angular dependence of the three-point correlation function of the
cosmic microwave background radiation as predicted by inflationary
cosmologies,''
Astrophys.\ J.\  {\bf 403}, L1 (1993)
[arXiv:astro-ph/9208001].

L.~M.~Wang and M.~Kamionkowski,
``The cosmic microwave background bispectrum and inflation,''
Phys.\ Rev.\  D {\bf 61}, 063504 (2000)
[arXiv:astro-ph/9907431].

L.~Verde, R.~Jimenez, M.~Kamionkowski and S.~Matarrese,
``Tests for primordial non-Gaussianity,''
Mon.\ Not.\ Roy.\ Astron.\ Soc.\  {\bf 325}, 412 (2001)
[arXiv:astro-ph/0011180].

\bibitem{loop_corrections}

S.~Weinberg,
``Quantum contributions to cosmological correlations,''
Phys.\ Rev.\  D {\bf 72}, 043514 (2005)
[arXiv:hep-th/0506236].

S.~Weinberg,
``Quantum contributions to cosmological correlations. II: Can these corrections become large?,''
Phys.\ Rev.\  D {\bf 74}, 023508 (2006)
[arXiv:hep-th/0605244].

D.~Seery,
``One-loop corrections to a scalar field during inflation,''
JCAP {\bf 0711}, 025 (2007)
[arXiv:0707.3377 [astro-ph]].
 
A.~Riotto and M.~S.~Sloth,
``On Resumming Inflationary Perturbations beyond One-loop,''
JCAP {\bf 0804}, 030 (2008)
[arXiv:0801.1845 [hep-ph]].

H.~R.~S.~Cogollo, Y.~Rodriguez and C.~A.~Valenzuela-Toledo,
``On the Issue of the ${\zeta}$ Series Convergence and Loop Corrections in the Generation of Observable Primordial Non-Gaussianity in Slow-Roll Inflation. Part I: the Bispectrum,''
JCAP {\bf 0808}, 029 (2008)
[arXiv:0806.1546 [astro-ph]].

 \bibitem{curvaton_ng}
 
D.~H.~Lyth, C.~Ungarelli and D.~Wands,
``The primordial density perturbation in the curvaton scenario,''
Phys.\ Rev.\  D {\bf 67}, 023503 (2003)
[arXiv:astro-ph/0208055].

N.~Bartolo, S.~Matarrese and A.~Riotto,
``On non-Gaussianity in the curvaton scenario,''
Phys.\ Rev.\  D {\bf 69}, 043503 (2004)
[arXiv:hep-ph/0309033].

 M.~Sasaki, J.~Valiviita and D.~Wands,
``Non-Gaussianity of the primordial perturbation in the curvaton model,''
Phys.\ Rev.\  D {\bf 74}, 103003 (2006)
[arXiv:astro-ph/0607627].

K.~A.~Malik and D.~H.~Lyth,
``A numerical study of non-Gaussianity in the curvaton scenario,''
JCAP {\bf 0609}, 008 (2006)
[arXiv:astro-ph/0604387].

H.~Assadullahi, J.~Valiviita and D.~Wands,
``Primordial non-Gaussianity from two curvaton decays,''
Phys.\ Rev.\  D {\bf 76}, 103003 (2007)
[arXiv:0708.0223 [hep-ph]].

Q.~G.~Huang,
``N-vaton,''
JCAP {\bf 0809}, 017 (2008)
[arXiv:0807.1567 [hep-th]].

\bibitem{single_field_ng}

P.~Creminelli,
``On non-Gaussianities in single-field inflation,''
JCAP {\bf 0310}, 003 (2003)
[arXiv:astro-ph/0306122].

X.~Chen, R.~Easther and E.~A.~Lim,
``Large non-Gaussianities in single field inflation,''
JCAP {\bf 0706}, 023 (2007)
[arXiv:astro-ph/0611645].

X.~Chen, R.~Easther and E.~A.~Lim,
``Generation and Characterization of Large Non-Gaussianities in Single Field
Inflation,''
JCAP {\bf 0804}, 010 (2008)
[arXiv:0801.3295 [astro-ph]].




\bibitem{inflation_ng}

N.~Bartolo, S.~Matarrese and A.~Riotto,
``Primordial non-Gaussianity from different cosmological scenarios,''
Nucl.\ Phys.\ Proc.\ Suppl.\  {\bf 148} (2005) 56.

D.~H.~Lyth and Y.~Rodriguez,
``The inflationary prediction for primordial non-Gaussianity,''
Phys.\ Rev.\ Lett.\  {\bf 95}, 121302 (2005)
[arXiv:astro-ph/0504045].

D.~H.~Lyth and Y.~Rodriguez,
``Non-Gaussianity from the second-order cosmological perturbation,''
Phys.\ Rev.\  D {\bf 71}, 123508 (2005)
[arXiv:astro-ph/0502578].

 F.~Bernardeau, T.~Brunier and J.~P.~Uzan,
``Models of inflation with primordial non-Gaussianities,''
AIP Conf.\ Proc.\  {\bf 861}, 821 (2006)
[arXiv:astro-ph/0604200].

I.~G.~Moss and C.~M.~Graham,
``Testing models of inflation with CMB non-Gaussianity,''
JCAP {\bf 0711}, 004 (2007)
[arXiv:0707.1647 [astro-ph]].

\bibitem{multiefield_ng}
K.~Enqvist and A.~Vaihkonen,
``Non-Gaussian perturbations in hybrid inflation,''
JCAP {\bf 0409}, 006 (2004)
[arXiv:hep-ph/0405103].

 L.~E.~Allen, S.~Gupta and D.~Wands,
``Non-Gaussian perturbations from multi-field inflation,''
JCAP {\bf 0601}, 006 (2006)
[arXiv:astro-ph/0509719].

F.~Vernizzi and D.~Wands,
``Non-Gaussianities in two-field inflation,''
JCAP {\bf 0605}, 019 (2006)
[arXiv:astro-ph/0603799].

 S.~A.~Kim and A.~R.~Liddle,
``Nflation: Multi-field inflationary dynamics and perturbations,''
Phys.\ Rev.\  D {\bf 74}, 023513 (2006)
[arXiv:astro-ph/0605604].

T.~Battefeld and R.~Easther,
``Non-Gaussianities in multi-field inflation,''
JCAP {\bf 0703}, 020 (2007)
[arXiv:astro-ph/0610296].

 D.~Battefeld and T.~Battefeld,
``Non-Gaussianities in N-flation,''
JCAP {\bf 0705}, 012 (2007)
[arXiv:hep-th/0703012].

N.~Barnaby and J.~M.~Cline,
``NonGaussianity from tachyonic preheating in hybrid inflation,''
Phys.\ Rev.\  D {\bf 75}, 086004 (2007)
[arXiv:astro-ph/0611750].

S.~Yokoyama, T.~Suyama and T.~Tanaka,
``Primordial Non-Gaussianity in Multi-Scalar Slow-Roll Inflation,''
Phys.\ Rev.\  D {\bf 77}, 083511 (2008)
[arXiv:0705.3178 [astro-ph]].

M.~x.~Huang, G.~Shiu and B.~Underwood,
``Multifield DBI Inflation and Non-Gaussianities,''
Phys.\ Rev.\  D {\bf 77}, 023511 (2008)
[arXiv:0709.3299 [hep-th]].

\bibitem{ekpyrosis_ng}
K.~Koyama, S.~Mizuno, F.~Vernizzi and D.~Wands,
``Non-Gaussianities from ekpyrotic collapse with multiple fields,''
JCAP {\bf 0711}, 024 (2007)
[arXiv:0708.4321 [hep-th]].

E.~I.~Buchbinder, J.~Khoury and B.~A.~Ovrut,
``Non-Gaussianities in New Ekpyrotic Cosmology,''
Phys.\ Rev.\ Lett.\  {\bf 100}, 171302 (2008)
[arXiv:0710.5172 [hep-th]].

J.~L.~Lehners and P.~J.~Steinhardt,
``Non-Gaussian Density Fluctuations from Entropically Generated Curvature
Perturbations in Ekpyrotic Models,''
Phys.\ Rev.\  D {\bf 77}, 063533 (2008)
[arXiv:0712.3779 [hep-th]].

\bibitem{Bartolo2004}
N.~Bartolo, E.~Komatsu, S.~Matarrese and A.~Riotto,
``Non-Gaussianity from inflation: Theory and observations,''
Phys.\ Rept.\  {\bf 402} (2004) 103
[arXiv:astro-ph/0406398]. 

\bibitem{Chen:2004}
X.~Chen,
``Multi-throat brane inflation,''
Phys.\ Rev.\  D {\bf 71} (2005) 063506
[arXiv:hep-th/0408084].

\bibitem{single_field_ng_2}
D.~Seery and J.~E.~Lidsey,
``Primordial non-Gaussianities in single field inflation,''
JCAP {\bf 0506}, 003 (2005)
[arXiv:astro-ph/0503692]. 

X.~Chen, M.~x.~Huang, S.~Kachru and G.~Shiu,
``Observational signatures and non-Gaussianities of general single field
inflation,''
JCAP {\bf 0701}, 002 (2007)
[arXiv:hep-th/0605045].


% DBI inflation

\bibitem{Silverstein:2003}
E.~Silverstein and D.~Tong,
``Scalar speed limits and cosmology: Acceleration from D-cceleration,''
Phys.\ Rev.\  D {\bf 70} (2004) 103505
[arXiv:hep-th/0310221].

\bibitem{Silverstein:2004}
M.~Alishahiha, E.~Silverstein and D.~Tong,
``DBI in the sky,''
Phys.\ Rev.\  D {\bf 70} (2004) 123505
[arXiv:hep-th/0404084].

\bibitem{Khoury:2008wj}
  J.~Khoury and F.~Piazza,
  ``Rapidly-Varying Speed of Sound, Scale Invariance and Non-Gaussian
  Signatures,''
  arXiv:0811.3633 [hep-th].


%TRISPECTRUM

\bibitem{trispectrum_ng}
C.~T.~Byrnes, M.~Sasaki and D.~Wands,
``The primordial trispectrum from inflation,''
Phys.\ Rev.\  D {\bf 74}, 123519 (2006)
[arXiv:astro-ph/0611075].

 M.~x.~Huang and G.~Shiu,
``The inflationary trispectrum for models with large non-Gaussianities,''
Phys.\ Rev.\  D {\bf 74}, 121301 (2006)
[arXiv:hep-th/0610235].

 D.~Seery and J.~E.~Lidsey,
``Non-Gaussianity from the inflationary trispectrum,''
JCAP {\bf 0701}, 008 (2007)
[arXiv:astro-ph/0611034].

F.~Arroja and K.~Koyama,
``Non-Gaussianity from the trispectrum in general single field inflation,''
Phys.\ Rev.\  D {\bf 77}, 083517 (2008)
[arXiv:0802.1167 [hep-th]].

%limits on detection

\bibitem{Creminelli2003}
P.~Creminelli, ``On non-gaussianities in single-field inflation,''
JCAP {\bf 0310} (2003) 003
[arXiv:astro-ph/0306122].

\bibitem{Creminelli2004}
P.~Creminelli and M.~Zaldarriaga,
``CMB 3-point functions generated by nonlinearities at recombination,''
Phys.\ Rev.\  D {\bf 70}, 083532 (2004)
[arXiv:astro-ph/0405428].

\bibitem{Serra2008}
P.~Serra and A.~Cooray,
``Impact of Secondary non-Gaussianities on the Search for Primordial
Non-Gaussianity with CMB Maps,''
arXiv:0801.3276 [astro-ph].

\bibitem{Komatsu2001a}
E.~Komatsu and D.~N.~Spergel,
``Acoustic signatures in the primary microwave background bispectrum,''
Phys.\ Rev.\  D {\bf 63}, 063002 (2001)
[arXiv:astro-ph/0005036].

\bibitem{Spergel2006}
D.~N.~Spergel {\it et al.}  [WMAP Collaboration],
``Wilkinson Microwave Anisotropy Probe (WMAP) three year results:
Implications for cosmology,''
Astrophys.\ J.\ Suppl.\  {\bf 170}, 377 (2007)
[arXiv:astro-ph/0603449].

\bibitem{Babich2004b}
D.~Babich and M.~Zaldarriaga,
``Primordial Bispectrum Information from CMB Polarization,''
Phys.\ Rev.\  D {\bf 70}, 083005 (2004)
[arXiv:astro-ph/0408455].

\bibitem{Riotto2002}
V.~Acquaviva, N.~Bartolo, S.~Matarrese and A.~Riotto,
``Second-order cosmological perturbations from inflation,''
Nucl.\ Phys.\  B {\bf 667} (2003) 119
[arXiv:astro-ph/0209156].

\bibitem{Maldacena2002}
J.~M.~Maldacena,
``Non-Gaussian features of primordial fluctuations in single field
inflationary models,''
JHEP {\bf 0305}, 013 (2003)
[arXiv:astro-ph/0210603].

\bibitem{Danielsson2002}
U.~H.~Danielsson,
``A note on inflation and transplanckian physics,''
Phys.\ Rev.\  D {\bf 66} (2002) 023511
[arXiv:hep-th/0203198].

\bibitem{Easther2002}
R.~Easther, B.~R.~Greene, W.~H.~Kinney and G.~Shiu,
``A generic estimate of trans-Planckian modifications to the primordial power
spectrum in inflation,''
Phys.\ Rev.\  D {\bf 66} (2002) 023518
[arXiv:hep-th/0204129].

\bibitem{SSS2004}
K.~Schalm, G.~Shiu and J.~P.~van der Schaar,
``Decoupling in an expanding universe: Boundary RG-flow affects initial conditions for inflation,''
JHEP {\bf 0404} (2004) 076
[arXiv:hep-th/0401164].

\bibitem{GSSS2005}
B.~Greene, K.~Schalm, J.~P.~van der Schaar and G.~Shiu,
``Extracting new physics from the CMB,''
{\it In the Proceedings of 22nd Texas Symposium on Relativistic Astrophysics at Stanford University, Stanford, California, 13-17 Dec 2004, pp
0001}
[arXiv:astro-ph/0503458].


%non-Gaussian detection:


\bibitem{Schmalzing1995}
  J.~Schmalzing, M.~Kerscher and T.~Buchert,
  ``Minkowski functionals in cosmology,''
  [arXiv:astro-ph/9508154].

\bibitem{Hobson1999}
  M.~Hobson, A.~Jones and A.~Lasenby,
  ``Wavelet analysis and the detection of non-Gaussianity in the CMB,''
  [arXiv:astro-ph/9810200].

\bibitem{Komatsu2003a}
E.~Komatsu {\it et al.}  [WMAP Collaboration],
``First Year Wilkinson Microwave Anisotropy Probe (WMAP) Observations: Tests
of Gaussianity,''
Astrophys.\ J.\ Suppl.\  {\bf 148}, 119 (2003)
[arXiv:astro-ph/0302223].

\bibitem{Creminelli2005}
P.~Creminelli, A.~Nicolis, L.~Senatore, M.~Tegmark and M.~Zaldarriaga,
``Limits on non-Gaussianities from WMAP data,''
JCAP {\bf 0605}, 004 (2006)
[arXiv:astro-ph/0509029].

\bibitem{CSZT2007}
P.~Creminelli, L.~Senatore, M.~Zaldarriaga and M.~Tegmark,
``Limits on $f_{NL}$ parameters from WMAP 3yr data,''
JCAP {\bf 0703}, 005 (2007)
[arXiv:astro-ph/0610600].

\bibitem{Yadav2007c}
A.~P.~S.~Yadav and B.~D.~Wandelt,
``Detection of primordial non-Gaussianity (fNL) in the WMAP 3-year data at
above 99.5 $\%$ confidence,''
arXiv:0712.1148 [astro-ph].

\bibitem{Komatsu2008}
E.~Komatsu {\it et al.}  [WMAP Collaboration],
``Five-Year Wilkinson Microwave Anisotropy Probe (WMAP)
Observations:Cosmological Interpretation,''
arXiv:0803.0547 [astro-ph].

\bibitem{Pyne1995}
T.~Pyne and S.~M.~Carroll,
``Higher-Order Gravitational Perturbations of the Cosmic Microwave
Background,''
Phys.\ Rev.\  D {\bf 53}, 2920 (1996)
[arXiv:astro-ph/9510041].


%shape+way to observe using templates    

\bibitem{Komatsu2003}
E.~Komatsu, D.~N.~Spergel and B.~D.~Wandelt,
``Measuring primordial non-Gaussianity in the cosmic microwave background,''
Astrophys.\ J.\  {\bf 634}, 14 (2005)
[arXiv:astro-ph/0305189].

\bibitem{Babich2004a}
D.~Babich, P.~Creminelli and M.~Zaldarriaga,
``The shape of non-Gaussianities,''
JCAP {\bf 0408}, 009 (2004)
[arXiv:astro-ph/0405356].

\bibitem{Liguori2005}
M.~Liguori, F.~K.~Hansen, E.~Komatsu, S.~Matarrese and A.~Riotto,
``Testing Primordial Non-Gaussianity in CMB Anisotropies,''
Phys.\ Rev.\  D {\bf 73}, 043505 (2006)
[arXiv:astro-ph/0509098].

\bibitem{Kogo2006}
N.~Kogo and E.~Komatsu,
``Angular Trispectrum of CMB Temperature Anisotropy from Primordial
Non-Gaussianity with the Full Radiation Transfer Function,''
Phys.\ Rev.\  D {\bf 73}, 083007 (2006)
[arXiv:astro-ph/0602099].

\bibitem{Creminelli2006}
P.~Creminelli, L.~Senatore and M.~Zaldarriaga,
``Estimators for local non-Gaussianities,''
JCAP {\bf 0703}, 019 (2007)
[arXiv:astro-ph/0606001].

\bibitem{Yadav2007a}
A.~P.~S.~Yadav, E.~Komatsu and B.~D.~Wandelt,
``Fast Estimator of Primordial Non-Gaussianity from Temperature and
Polarization Anisotropies in the Cosmic Microwave Background,''
Astrophys.\ J.\  {\bf 664}, 680 (2007)
[arXiv:astro-ph/0701921].

\bibitem{Liguori2007}
M.~Liguori, A.~Yadav, F.~K.~Hansen, E.~Komatsu, S.~Matarrese and B.~Wandelt,
``Temperature and Polarization CMB Maps from Primordial non-Gaussianities of
the Local Type,''
Phys.\ Rev.\  D {\bf 76}, 105016 (2007)
[Erratum-ibid.\  D {\bf 77}, 029902 (2008)]
[arXiv:0708.3786 [astro-ph]].

\bibitem{Yadav2007b}
A.~P.~S.~Yadav, E.~Komatsu, B.~D.~Wandelt, M.~Liguori, F.~K.~Hansen and S.~Matarrese,
``Fast Estimator of Primordial Non-Gaussianity from Temperature and
Polarization Anisotropies in the Cosmic Microwave Background II: Partial Sky
Coverage and Inhomogeneous Noise,''
arXiv:0711.4933 [astro-ph].


%WMAP observations:

\bibitem{Spergel2003}
D.~N.~Spergel {\it et al.}  [WMAP Collaboration],
``First Year Wilkinson Microwave Anisotropy Probe (WMAP) Observations:
Determination of Cosmological Parameters,''
Astrophys.\ J.\ Suppl.\  {\bf 148}, 175 (2003)
[arXiv:astro-ph/0302209].

\bibitem{Peiris2003}
H.~V.~Peiris {\it et al.},
``First year Wilkinson Microwave Anisotropy Probe (WMAP) observations:
Implications for inflation,''
Astrophys.\ J.\ Suppl.\  {\bf 148}, 213 (2003)
[arXiv:astro-ph/0302225].



%INITIAL STATE MOD.

\bibitem{Martin1999}
J.~Martin, A.~Riazuelo and M.~Sakellariadou,
``Non-vacuum initial states for cosmological perturbations of
quantum-mechanical origin,''
Phys.\ Rev.\  D {\bf 61}, 083518 (2000)
[arXiv:astro-ph/9904167].

\bibitem{Gangui2002}
A.~Gangui, J.~Martin and M.~Sakellariadou,
``Single field inflation and non-Gaussianity,''
Phys.\ Rev.\  D {\bf 66}, 083502 (2002)
[arXiv:astro-ph/0205202].

\bibitem{Porrati2004a}
M.~Porrati,
``Bounds on generic high-energy physics modifications to the primordial
power spectrum from back-reaction on the metric,''
Phys.\ Lett.\  B {\bf 596}, 306 (2004)
[arXiv:hep-th/0402038].

\bibitem{Porrati2004b}
M.~Porrati,
``Effective field theory approach to cosmological initial conditions:
Self-consistency bounds and non-Gaussianities,''
arXiv:hep-th/0409210.

\bibitem{Shiu2006}
X.~Chen, M.~x.~Huang, S.~Kachru and G.~Shiu,
``Observational signatures and non-Gaussianities of general single field
inflation'',
JCAP {\bf 0701} (2007) 002
[arXiv:hep-th/0605045].


\bibitem{Holman2007}
R.~Holman and A.~J.~Tolley,
``Enhanced Non-Gaussianity from Excited Initial States'',
JCAP {\bf 0805}, 001 (2008)
[arXiv:0710.1302 [hep-th]].

\bibitem{Meerburg2008}
P. D. ~Meerburg and J.~P. ~van~der~Schaar,
``Non-Gaussianities from Modified Initial State'',
work in preparation.

\bibitem{Fergusson2006}
J.~R.~Fergusson and E.~P.~S.~Shellard,
``Primordial non-Gaussianity and the CMB bispectrum,''
Phys.\ Rev.\  D {\bf 76} (2007) 083523
[arXiv:astro-ph/0612713].


\bibitem{Fergusson2008}
J.~R.~Fergusson and E.~P.~S.~Shellard,
``The shape of primordial non-Gaussianity and the CMB bispectrum,''
arXiv:0812.3413 [astro-ph].

\bibitem{Smith2006}
K.~M.~Smith and M.~Zaldarriaga,
``Algorithms for bispectra: forecasting, optimal analysis, and simulation'',
arXiv:astro-ph/0612571.

\bibitem{Zaldarriaga}
U. Seljak and M. Zaldarriaga,
``A line of sight approach to Cosmic Microwave Background anisotropies'',
Astrophys. J., \textbf{469}, 437 (1996)

%\FergussonRA


\end{thebibliography}
\end{document}